\RequirePackage{lineno}
\documentclass[12pt,onecolumn,superscriptaddress,amsmath,amssymb]{revtex4}

\usepackage{times, color}
\usepackage{graphicx,amsmath,amssymb}
\usepackage{dcolumn}
\usepackage{bm}
\usepackage[printwatermark]{xwatermark}
\usepackage{xcolor}
\usepackage{lipsum}
\usepackage{verbatim}
\usepackage{epstopdf}
\usepackage{lineno}
\usepackage[colorlinks=true, urlcolor=blue, linkcolor=blue, citecolor=blue]{hyperref}
\usepackage{xfrac}
\usepackage{ulem}

\begin{document} 

\title{Measurement of the generalized spin polarizabilities of the neutron in the low $Q^2$ region}

\author{Vincent~Sulkosky}
\affiliation{William \& Mary, Williamsburg, Virginia 23187-8795, USA }
\affiliation{Thomas Jefferson National Accelerator Facility, Newport News, Virginia 23606, USA }
\affiliation{University of Virginia, Charlottesville, Virginia 22904, USA}
\author{Chao Peng}
\affiliation{Duke University, Durham, North Carolina 27708, USA}
\affiliation{Argonne National Laboratory, Lemont, Illinois 60439, USA}
\author{Jian-ping Chen}
\affiliation{Thomas Jefferson National Accelerator Facility, Newport News, Virginia 23606, USA }
\author{Alexandre Deur\footnote{Corresponding author; E-mail:  deurpam@jlab.org.}}
\affiliation{University of Virginia, Charlottesville, Virginia 22904, USA}
\affiliation{Thomas Jefferson National Accelerator Facility, Newport News, Virginia 23606, USA }
\author{Sergey Abrahamyan}
\affiliation{Yerevan Physics Institute, Yerevan 375036, Armenia}
\author{Konrad~A. Aniol}
\affiliation{California State University, Los Angeles, Los Angeles, California 90032, USA}
\author{David S. Armstrong}
\affiliation{William \& Mary, Williamsburg, Virginia 23187-8795, USA }
\author{Todd Averett}
\affiliation{William \& Mary, Williamsburg, Virginia 23187-8795, USA }
\author{Stephanie L. Bailey}
\affiliation{William \& Mary, Williamsburg, Virginia 23187-8795, USA }
\author{Arie Beck}
\affiliation{Massachusetts Institute of Technology, Cambridge, Massachusetts 02139, USA}
% ...
\collaboration{Jefferson Lab E97-110 Collaboration}

\date{\today}

\baselineskip24pt

\begin{abstract} 
Understanding the nucleon spin structure in the regime where the strong interaction becomes truly strong poses a 
challenge to both experiment and theory. At energy scales below the nucleon mass of about 1~GeV, the intense 
interaction among the quarks and gluons inside the nucleon makes them highly correlated. 
Their coherent behaviour causes the emergence of effective degrees of freedom, requiring the application of 
non-perturbative techniques, such as chiral effective field theory~\cite{Bernard:1995dp}. 
Here, we present measurements of the neutron's generalized spin-polarizabilities that quantify the neutron's 
spin precession under electromagnetic fields at very low energy-momentum transfer squared down to 0.035~GeV$^2$. 
In this regime, chiral effective field theory calculations~\cite{Bernard:2012hb, Lensky:2014dda, Alarcon:2020icz} 
are expected to be applicable. Our data, however, show a strong discrepancy with these predictions, 
presenting a challenge to the current description of the neutron's spin properties.

\end{abstract}

\maketitle

\begin{figure}[!h]
  \centering
    \includegraphics[width=0.35\textwidth]{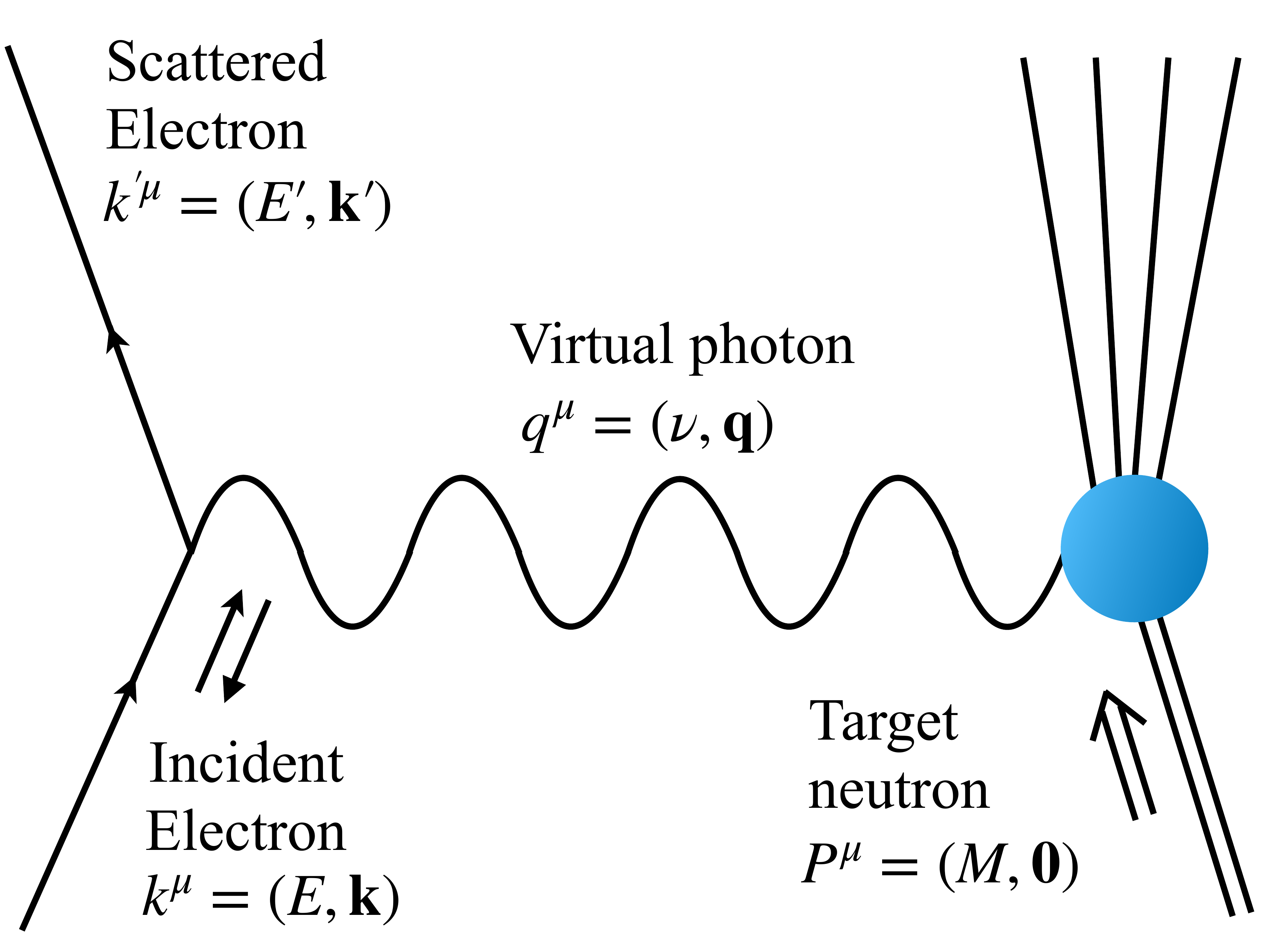}
 \caption{\label{Fig:eN_born1app} 
Electron scattering off a neutron by the one-photon exchange process. 
  The 4-momenta of the incident and the scattered electrons are 
 $k^\mu =(E,\mathbf{k})$ and $k'^\mu =(E',\mathbf{k'})$, respectively,
 and that of the photon is $q^\mu =(\nu,\mathbf{q})$. The neutron, at rest in the laboratory frame, 
 has a 4-momentum $P^\mu=(M,\mathbf{0})$
 The arrows $\uparrow\downarrow$ represent the spin direction of the incident electron and 
 $\Uparrow$ that of the neutron. 
 The generalized spin-polarizabilities of the neutron can be measured when 
 both the incident electron and the
neutron are polarized.}
\end{figure}

The nucleon is the basic building block of nature, 
accounting for about 99\% of the universe's visible mass. Understanding its
properties, e.g.,   mass and spin,  is thus crucial. Those are mainly determined by the Strong Interaction, 
which is described by Quantum Chromodynamics (QCD) with quarks and gluons as the fundamental degrees of freedom. 
The nucleon structure is satisfactorily understood at  high $Q^2$ (short space-time scales,  see Fig.~\ref{Fig:eN_born1app}
for the definition of kinematic variables),
since there QCD is calculable using perturbation methods (perturbative QCD) 
and tested by numerous experimental measurements. 
At lower $Q^2$,  the strong coupling $\alpha_s$ becomes too large 
for perturbative QCD to be applicable~\cite{Deur:2016tte}. 
Yet, calculations are critically needed 
since  the Strong Interaction's chiral symmetry breaks in this region. 
Chiral symmetry and its breaking is one of the most important properties of the Strong Interaction and
is believed to lead to the emergence of the nucleon's global properties.
To understand  how the underlying structure leads to the emergence of these global properties, non-perturbative methods 
must be used. A method using the fundamental quark and gluon degrees of freedom is 
lattice QCD. However, calculations from this method are often intractable for spin observables at low $Q^2$~\cite{Deur:2018roz}. 
Another solution is to employ effective theories.  Chiral effective field theory ($\chi$EFT) capitalizes on 
QCD's approximate chiral symmetry and uses the emergent hadronic degrees of freedom. Therein 
lies $\chi$EFT's strengths and challenges: 
 while the nucleon and the pion are used for first-order calculations, this is often  
insufficient to describe the data, and
heavier hadrons, such as the nucleon's first excited state $\Delta(1232)$, 
  become needed. This complicates  $\chi$EFT calculations,
and theorists are still seeking the best way to include the $\Delta(1232)$ in their calculations.
It is therefore crucial to perform precision measurements  at  low enough $Q^2$
to test $\chi$EFT calculations.    Spin observables, among them the generalized 
spin-polarizabilities that are reported here,
 provide an extensive set of tests to benchmark $\chi$EFT calculations~\cite{Deur:2018roz}.

Polarizabilities describe how the components of an object collectively react to external electromagnetic fields. 
In particular, spin-polarizabilities quantify the object's spin precession under an electromagnetic field.
The spin-polarizabilities, initially defined with real photons, can be generalized to virtual photons
such as those used to probe the neutron in our experiment.
Accordingly, generalized spin-polarizabilities are extracted by scattering polarized electrons off polarized nucleons 
and measuring how the cross-section changes when the relative orientation between the electron and 
nucleon spins is varied (see Fig.~\ref{Fig:eN_born1app}). 
The energy-momentum transferred between the electron and neutron is $(\nu,\bm{q})$, 
with $Q^{2}=\bm{q}^2-\nu^2$ characterizing the space-time scale at which we probe the neutron. 
While real photons ($Q^2 = 0$) only have transverse polarizations, mediating virtual 
photons  ($Q^2\neq 0$) are transversely ($\mathrm{T}$) or longitudinally ($\mathrm{L}$) polarized. 
Thus, 
two contributions to the spin-polarizability arise: one from the transverse-transverse ($\mathrm{TT}$)
interference called the forward spin-polarizability $\gamma_0(Q^2)$, and the other from 
the longitudinal-transverse ($\mathrm{LT}$) interference, called the Longitudinal-Transverse interference polarizability 
$\delta_\mathrm{LT}(Q^2)$, which is available only with virtual photons. 
The additional  longitudinal polarization direction and the ensuing interference term offer extra 
latitude to test theories describing the Strong Interaction.

\begin{figure}[!h]
  \includegraphics[width=1\textwidth]{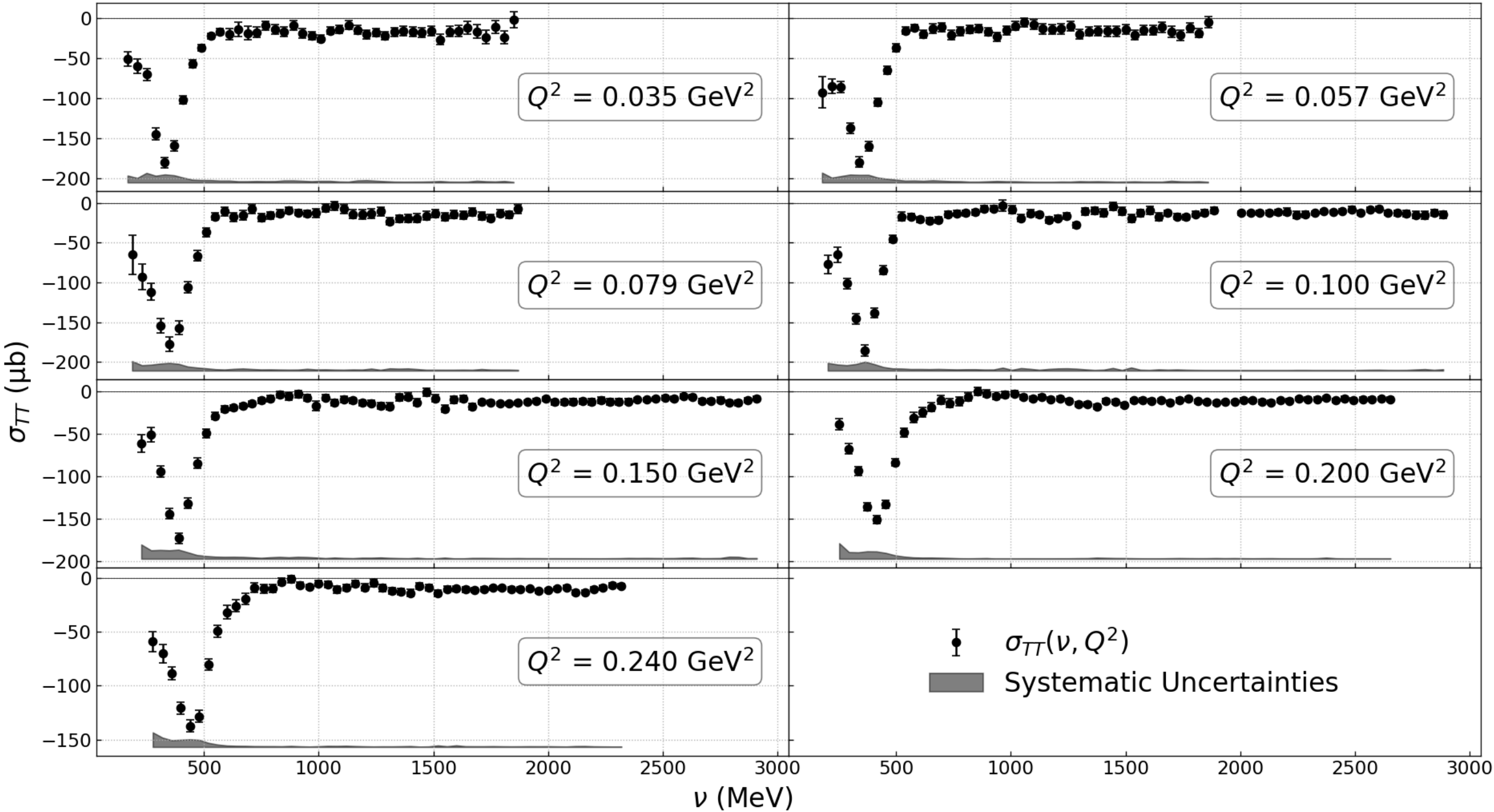} 
 \caption{\label{Fig:sigTT}  The transverse-transverse cross-section $\sigma_\mathrm{TT}(\nu,Q^2)$ 
 for $^3$He.  The data are displayed at the $Q^2$ values at which they are integrated 
to form $\gamma_0$ (Eq.~\ref{eq:gamma_0_SSF}). 
 The error bars, sometimes too small to be visible, represent the statistical uncertainties.
The systematic uncertainty is indicated by the band at the bottom of each panel.
The nuclear corrections providing the neutron information from the $^3$He data are applied after the integration. The prominent negative peak at small-$\nu$ is the 
$\Delta(1232)$ contribution. }
\end{figure}

\begin{figure}[!h]
  \includegraphics[width=1.\textwidth]{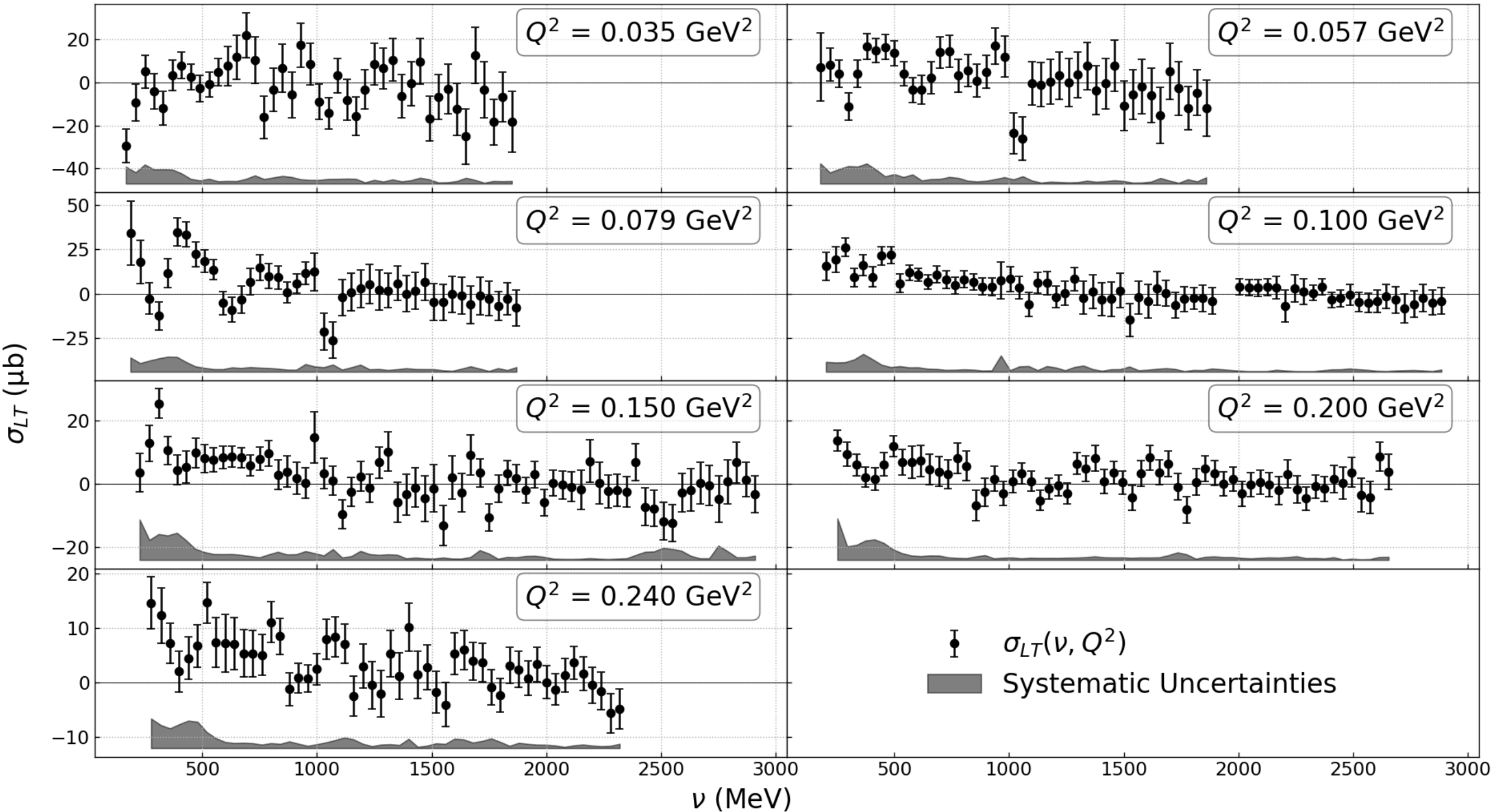} 
\caption{\label{sigma_LT} The longitudinal-transverse interference cross-section $\sigma_\mathrm{LT}(\nu,Q^2)$ 
for $^3$He.  The data are displayed at the $Q^2$ values at which they are integrated into $\delta_\mathrm{LT}(Q^2)$ (Eq.~\ref{eq:delta_{LT} SR})
or $I_\mathrm{LT}(Q^2)$ (Eq.~\ref{Eq.:SchwingerSR}). 
 The error bars represent the statistical uncertainties.
The systematic uncertainty is indicated by the band at the bottom of each panel.
The nuclear corrections~\cite{CiofidegliAtti:1996cg} necessary to obtain the neutron information from the $^3$He data are applied after the integration. The prominent $\Delta(1232)$ contribution 
seen for $\sigma_\mathrm{TT}(\nu,Q^2)$ in Fig.~\ref{Fig:sigTT} is not present here, in agreement with the expectation that the role of $\Delta(1232)$ is suppressed in $\mathrm{LT}$-interference quantities. }
\end{figure}

The theoretical basis to measure  $\delta_\mathrm{LT}(Q^2)$ originates from a work of Gell-Mann, 
Goldberger and Thirring~\cite{GellMann:1954db, Guichon:1995pu}. This work led to relations between 
the cross-sections measured in polarized electron-nucleon scattering  (Fig.~\ref{Fig:eN_born1app}) 
and the spin-polarizabilities:
\begin{eqnarray}
\gamma_0(Q^2) =  
\frac{1}{2\pi^2}\int_{\nu_0}^{\infty}\frac{\kappa_\gamma}{\nu^{2}}\frac{\sigma_\mathrm{TT}(\nu,Q^2)}{\nu^{2}}d\nu,
\label{eq:gamma_0_SSF}
\end{eqnarray}
 \begin{eqnarray}
\delta_\mathrm{LT}(Q^2) = \left(\frac{1}{2\pi^2}\right)\int_{\nu_0}^{\infty}\frac{\kappa_{\gamma}}{\nu Q}
\frac{\sigma_\mathrm{LT}(\nu,Q^2)}{\nu^2}d\nu ,
\label{eq:delta_{LT} SR}
\end{eqnarray}
where $\kappa_\gamma= \nu-\sfrac{Q^2}{2M}$~\cite{Hand:1963bb} is the photon flux factor, 
$\nu_0$ the photoproduction threshold,  
and $\sigma_\mathrm{TT}$ and $\sigma_\mathrm{LT}$ are respectively the 
$\mathrm{TT}$ and $\mathrm{LT}$ interference cross-sections.
They are obtained from~\cite{Deur:2018roz,Chen:2010qc}:
\begin{eqnarray}
\sigma_{TT}(\nu,Q^2) = \frac{\pi^2EQ^2(1 - \epsilon)}{\alpha\kappa_{\gamma}E'(1 - \epsilon E'/E)(1 + \eta\zeta)}\left(\sqrt{\frac{2\epsilon}{1+\epsilon}}\Delta\sigma_{\parallel}(\nu,Q^2) - \eta\Delta\sigma_{\perp}(\nu,Q^2)\right),  \label{eq:d sigma_TT} \\
\sigma_{LT}(\nu,Q^2) = \frac{\pi^2EQ^2(1 - \epsilon)}{\alpha\kappa_{\gamma}E'(1 - \epsilon E'/E)(1 + \eta\zeta)}\left(\sqrt{\frac{2\epsilon}{1+\epsilon}}\zeta\Delta\sigma_{\parallel}(\nu,Q^2) + \Delta\sigma_{\perp}(\nu,Q^2)\right), \label{eq:d sigma_LT}
\end{eqnarray}
where $\Delta \sigma_{||}$ ($\Delta \sigma_{\bot}$) is the difference between the cross sections 
when the beam and target spin directions are parallel and antiparallel (perpendicular),
$\alpha$ is the electromagnetic coupling constant, 
$\epsilon =1/[1+2(1+\sfrac{Q^2}{4M^2x^2})\tan^2(\sfrac{\theta}{2})]$ with $x= \sfrac{Q^2}{2m\nu}$ the  Bjorken scaling variable and $\theta$ the electron scattering angle in the laboratory frame,
$\eta = \sfrac{\epsilon Q}{(E-E'\epsilon)}$ and $\zeta = \sfrac{\eta(1 + \epsilon)}{2\epsilon}$. 
The $\sigma_\mathrm{TT}$ and  $\sigma_\mathrm{LT}$, shown in Figs.~\ref{Fig:sigTT} and~\ref{sigma_LT},
were integrated according to Eqs.~(\ref{eq:gamma_0_SSF}) and (\ref{eq:delta_{LT} SR})
to obtain  $\gamma_0(Q^2)$ and $\delta_\mathrm{LT}(Q^2)$. 
The unmeasured part of the integrals at large $\nu$  is often negligible due to the $\nu$-weighting. 

An outstanding feature of $\delta_\mathrm{LT}(Q^2)$  at low $Q^2$ is that the $\Delta(1232)$ 
is not expected to appreciably  contribute to   the LT-interference cross section, 
since exciting the $\Delta(1232)$ overwhelmingly involves transverse photons.
This should alleviate the difficulty of including the $\Delta(1232)$ in $\chi$EFT calculations, making them
more robust.
However, the first measurement of $\delta_\mathrm{LT}(Q^2)$ from JLab experiment  E94-010~\cite{Amarian:2004yf}
done at $Q^2 \geq 0.1$ GeV$^2$ strongly disagreed with $\chi$EFT calculations~\cite{Bernard:2002pw, Kao:2002cp}.
This surprising result, known as the ``$\delta_\mathrm{LT}$ puzzle''~\cite{Chen:2010qc}, triggered improved $\chi$EFT 
calculations~\cite{Hagelstein:2015egb} which now explicitly include
the $\Delta(1232)$~\cite{Bernard:2012hb, Lensky:2014dda, Alarcon:2020icz}, and measurements of $\delta_\mathrm{LT}$ at lower $Q^2$ 
where $\chi$EFT can be best tested. New data of $\delta_\mathrm{LT}$ on the neutron at very low $Q^2$
are presented next, which were taken during experiment JLab E97-110.

Eq.~(\ref{eq:delta_{LT} SR}) allows measuring $\delta_\mathrm{LT}^n(Q^2)$  
(the superscript $n$ indicates neutron quantities)
by scattering polarized electrons off polarized neutrons in $^3$He nuclei. 
The data were acquired in Hall A~\cite{Alcorn:2004sb} of Jefferson Lab (JLab) during experiment 
E97-110~\cite{Sulkosky:2019zmn}. 
The probing virtual photons were produced by a longitudinally
polarized electron beam during its scattering off a polarized $^{3}$He target~\cite{Alcorn:2004sb}. 
The beam polarization, flipped pseudo-randomly at 30 Hz and 
monitored by M\o ller and Compton polarimeters, was (75.0~$\pm$~2.3)\%.  
The beam energies ranged from 1.1 to 4.4 GeV, and the beam current was typically a few $\mu$A.
Since free neutrons are unstable, we used $^3$He nuclei as an effective polarized neutron target. To  first-order,
polarized $^3$He nuclei  can be treated as effective polarized neutrons together with unpolarized protons because the $^3$He's nucleons 
(two protons and one neutron) are mostly in an $S$-state, and so the Pauli exclusion principle
dictates that   in the S-state the proton spins point oppositely, yielding no net contribution to the $^3$He spin. 
The gaseous ($\approx12$~atm) $^{3}$He was contained in a 40~cm-long glass cylinder and   
polarized by spin-exchange optical pumping of Rubidium atoms. Helmholtz coils provided a longitudinal or transverse 2.5~mT 
field used to maintain the polarization, to orient it longitudinally or transversely (in-plane)
to the beam direction, and to aid in performing polarimetry. The average target polarization in-beam was (39.0~$\pm$~1.6)\%.
The scattered electrons from the reaction $\vec{\rm{^{3}He}}$($\vec{\rm{e}},\rm{e'}$) 
were detected by a High Resolution Spectrometer (HRS)~\cite{Alcorn:2004sb} 
supplemented by a dipole magnet~\cite{septum} allowing 
 us to detect electrons scattered at
angles down to 6$^{\circ}$. Behind the HRS, drift chambers provided
particle tracking, scintillator planes enabled the data acquisition trigger,
and a gas Cherenkov counter and electromagnetic calorimeters ensured the identification of the particle type.  

 The measured $\sigma_\mathrm{TT}$ ($\sigma_\mathrm{LT}$) on $^3$He is shown in Fig.~\ref{Fig:sigTT} (Fig.~\ref{sigma_LT}). Its 
values with their uncertainties are available in the Supplementary Data Files.
While polarized $^{3}$He nuclei are effectively polarized neutrons to good approximation, nuclear corrections 
are needed to obtain genuine neutron information. 
The prescription of Ref.~\cite{CiofidegliAtti:1996cg} was used for the correction. 
The effect of the nuclear correction, which can be obtained from Tables  I-III 
in the Supplementary Data Files, 
is relatively small. In particular it does not appreciably affect the $Q^2$ trend seen for the uncorrected $^3$He integrals.
The relative uncertainty on this correction is  estimated to be 6 to 14\%  relative 
to the correction, 
the higher uncertainties corresponding to our lowest $Q^2$ values.
The quasi-elastic contamination was corrected following the procedure described in~\cite{Sulkosky:2019zmn}. 
The correction is small for $\delta_\mathrm{LT}^n$, but important for $\gamma_0^n$ and was estimated 
using~\cite{Deltuva:2005wx}. No calculation uncertainty is provided in \cite{Deltuva:2005wx}  and using another 
quasi-elastic calculation~\cite{Golak:2005xq} may shift the lowest-$Q^2$ $\gamma_0^n$ data points by as 
much as our total systematic uncertainty.
The other main systematic uncertainties come from the absolute
cross-sections (3.5 to 4.5\%), target and beam polarizations (3 to 5\% and 3.5\%, respectively), and
radiative corrections (3 to 7\%). 

\begin{figure}[!h]
  \centering
    \includegraphics[width=1\textwidth]{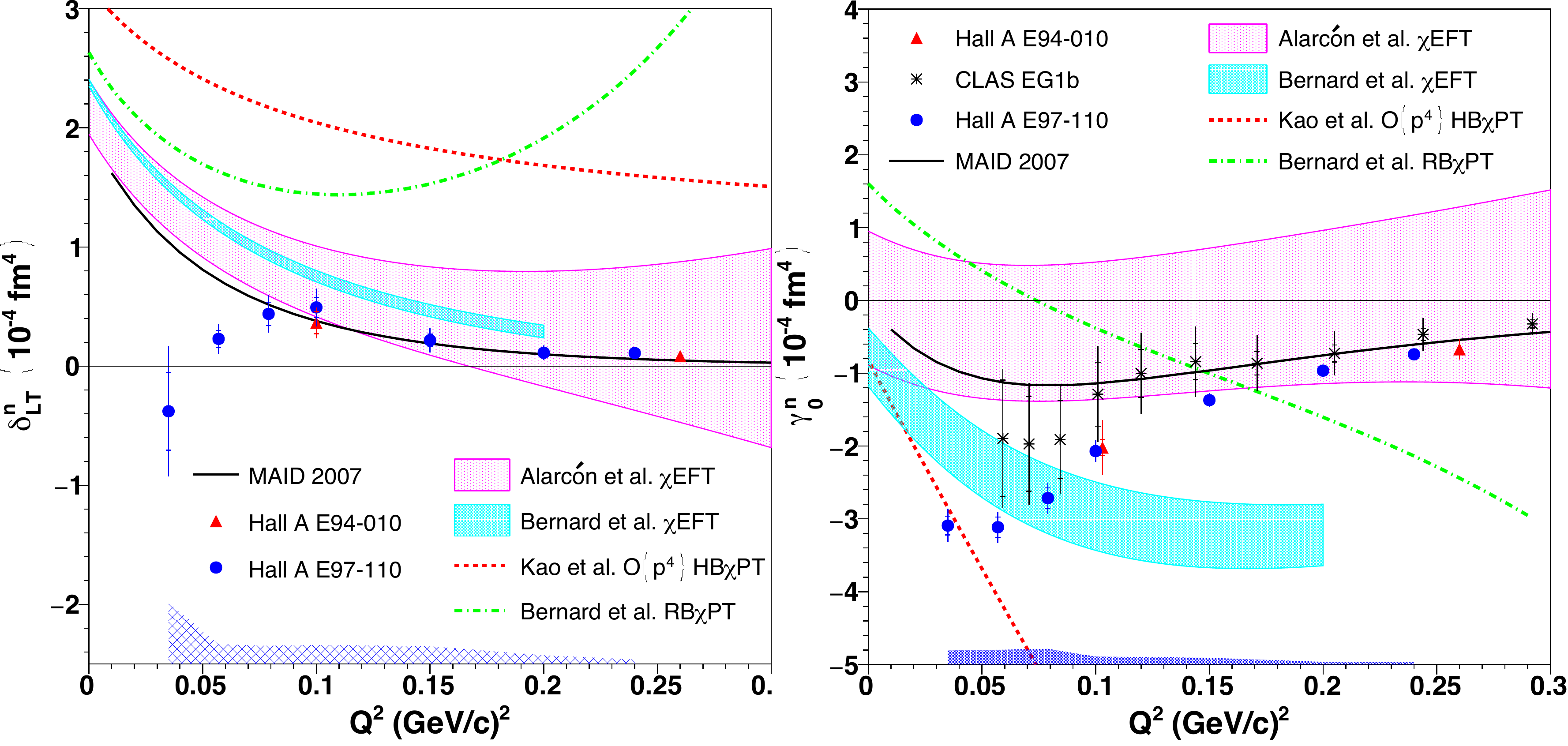}
 \caption{\label{Fig:delta_LT+gamma0} 
 The generalized spin polarizabilities $\delta_\mathrm{LT}^n(Q^2)$ and $\gamma_0^n(Q^2)$.
 Left:  The generalized spin-polarizability $\delta_\mathrm{LT}^n(Q^2)$. The circles represent the results
 from experiment E97-110.  They can be
 compared to earlier E94-010 data~\cite{Amarian:2004yf}  (triangles) and theoretical calculations: the
older $\chi$EFT calculations~\cite{Bernard:2002pw} (dot-dashed line) and~\cite{Kao:2002cp} (dashed line) in 
which the $\Delta$ resonance contribution is not included or included phenomenologically, the state-of-the-art calculations 
\cite{Bernard:2012hb} (cyan band) and \cite{Alarcon:2020icz} (magenta band)
that include the $\Delta$, as well as the MAID model~\cite{Drechsel:1998hk} (black curve) 
which is a fit to world resonance data.
For the E97-110 data, the inner error bars, sometimes too small to be visible, represent the statistical uncertainties. 
The outer error bars show the  statistical  and uncorrelated systematic uncertainties. 
The correlated systematic uncertainty is indicated by the band at the bottom.
 For the other experimental data, the error bars show the statistical and systematic uncertainties added in quadrature. 
 Right:   The generalized forward spin-polarizability $\gamma_0^n(Q^2)$,  using the same symbols 
as in the left panel. The asterisks represent the CLAS  data~\cite{Guler:2015hsw}.}
\end{figure}

Our  $\delta_\mathrm{LT}^n(Q^2)$ data are shown in the left panel of Fig.~\ref{Fig:delta_LT+gamma0}. They agree with earlier data from E94-010 at larger $Q^2$~\cite{Amarian:2004yf}
while reaching much lower $Q^2$  where the $\chi$EFT is expected to work well.
The measurement can be compared to $\chi$EFT calculations~\cite{Bernard:2002pw, Kao:2002cp, Bernard:2012hb, Alarcon:2020icz} 
and a model parameterization of the world photo- and electro-production data called MAID~\cite{Drechsel:1998hk}.
Earlier $\chi$EFT calculations~\cite{Bernard:2002pw, Kao:2002cp} used different approaches (Heavy Baryon and Relativistic Baryon chiral perturbation theory: 
HB$\chi$PT and RB$\chi$PT, respectively), and furthermore either neglected the $\Delta(1232)$ degrees of freedom,
or included it  approximately. Newer calculations~\cite{Bernard:2012hb, Lensky:2014dda, Alarcon:2020icz}, which are all fully relativistic,
account for the $\Delta(1232)$  explicitly by using a perturbative expansion, 
but they differ in their choice of expansion parameter.
Despite this theoretical improvement and the small-$Q^2$ reach that places our data well in the validity domain of $\chi$EFT, 
our $\delta_\mathrm{LT}^n(Q^2)$ starkly disagrees with the predictions.  
This is even more surprising because the latest $\chi$EFT calculations  of $\delta_\mathrm{LT}^n$
agree with each other, suggesting that  calculations for this particular observable should be under control.  
However, our data reveal an opposite trend  with $Q^2$ to that of all the $\chi$EFT calculations.

\begin{figure}[!h]
  \includegraphics[width=0.7\textwidth]{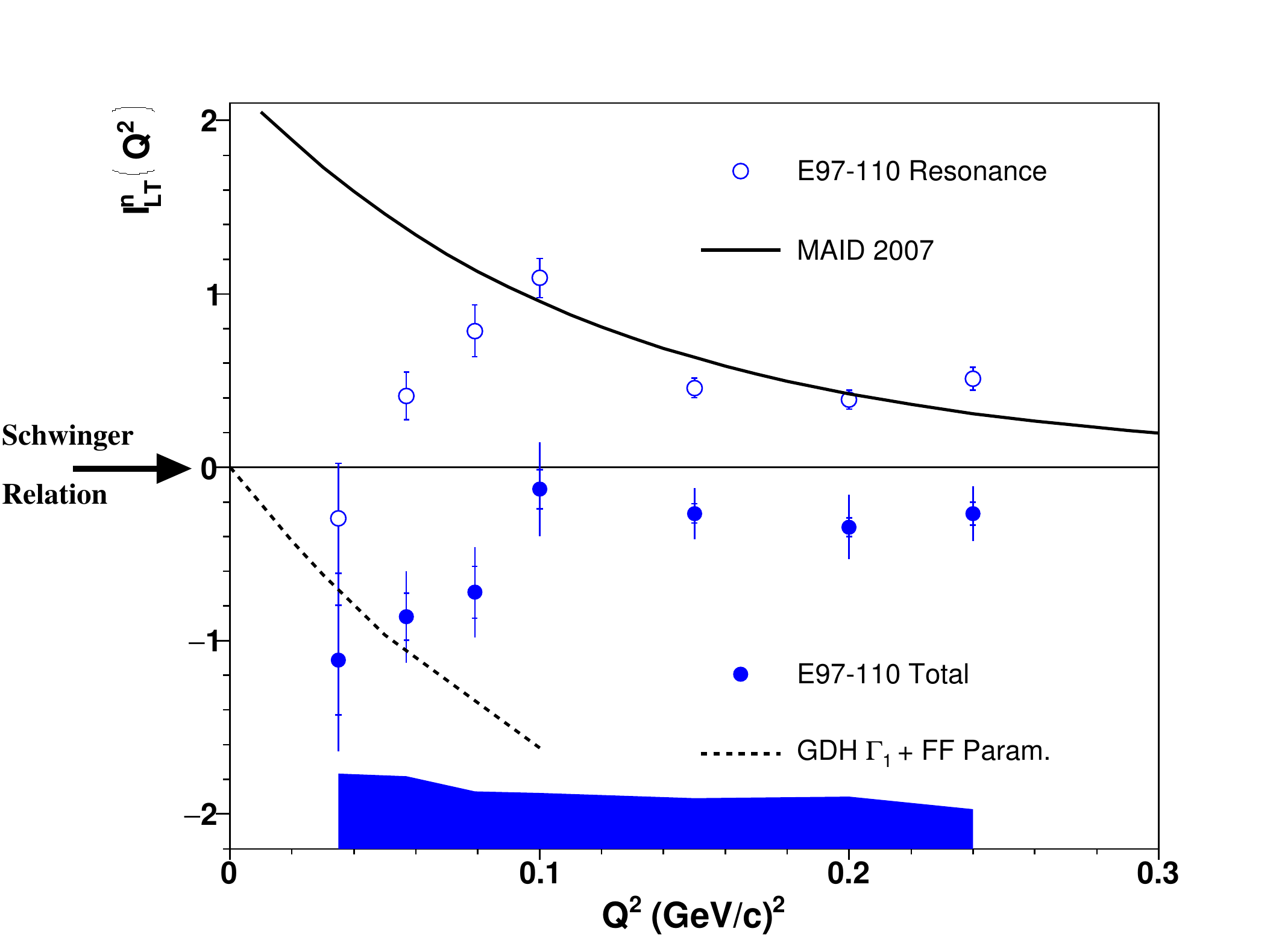} 
\caption{\label{SchwingerSR}  The   Schwinger integral $I_\mathrm{LT}^n(Q^2)$. 
The open symbols are our results without the large 
$\nu$ part of $I_\mathrm{LT}$.
The filled blue circles are our results for the full $I_\mathrm{LT}$, using an estimate for the large $\nu$ contribution.
The inner error bars represent the statistical uncertainties. 
The outer error bars show the combined statistical  and uncorrelated systematic uncertainties. 
The correlated systematic uncertainty is indicated by the band.
The Schwinger relation~\cite{Schwinger:1975ti} for the neutron predicts $I_\mathrm{LT}^n(0)=0$ at $Q^2=0$.
The plain line shows the MAID model~\cite{Drechsel:1998hk},  which is a fit to world resonance data (to be compared to the open symbols).
The dashed line uses the GDH  ($\Gamma_1$)~\cite{Gerasimov:1965et, Drell:1966jv} and Burkhardt-Cottingham~\cite{Burkhardt:1970ti} relations, together with an elastic form factor  (FF) parameterization~\cite{Ye:2017gyb}, to 
obtain $I_\mathrm{LT}^n(Q^2)$ for $Q^2 \to 0$.}
\end{figure}

This startling discrepancy demanded further scrutinization of our data. 
They are compatible with the E94-010 data where they overlap. 
This is also true for $\gamma_0^n(Q^2)$,  
which we measured concurrently and show in  the right panel of Fig.~\ref{Fig:delta_LT+gamma0}. 
The measured $\gamma_0^n(Q^2)$ also agrees 
with data from CLAS experiment EG1~\cite{Guler:2015hsw}, which 
used a target and detectors that are very different from  E97-110 and E94-010. 
Our $\gamma_0^n(Q^2)$ data  generally disagree with $\chi$EFT calculations. Since
$\gamma_0(Q^2)$ does not benefit from the suppression of the $\Delta(1232)$ contribution,
and since $\gamma^n_0(Q^2)$ predictions do not reach a consensus, 
this disagreement is not entirely surprising, in contrast to the unexpected $\delta_\mathrm{LT}^n(Q^2)$ disagreement.
Interestingly, we can also study with our data the Schwinger relation~\cite{Schwinger:1975ti},
 which has a similar definition but without $\nu^{-2}$ weighting in its integrand:
\begin{eqnarray}
I_\mathrm{LT}(Q^2) & \equiv & \left(\frac{M^2}{\alpha \pi^2}\right)\int_{\nu_0}^{\infty} \Bigl[\kappa_{\gamma} \frac{\sigma_\mathrm{LT}(\nu,Q^2)}{Q\nu}\Bigr]_{Q=0} d\nu.
\label{Eq.:SchwingerSR}
\end{eqnarray}
Schwinger predicted that $I_\mathrm{LT}(Q^2)  \xrightarrow [Q^2 \to 0]~ \kappa e_t $,
with $\kappa$ the anomalous magnetic moment of the target particle and $e_t$ its electric charge.
This prediction is general, e.g. it does not use $\chi$EFT.
$I_\mathrm{LT}(Q^2)$  having no $\sfrac{1}{\nu^2}$-weighting, 
the large $\nu$ contribution to the integral is not negligible. 
Since  this contribution to the integral cannot be measured,  a parameterization based on the model described in~\cite{Adhikari:2017wox} completed by a Regge-based parameterization~\cite{Bass:2018uon} for the largest $\nu$ part was used to extrapolate it.
Our measurement of $I^n_\mathrm{LT}(Q^2)$ is shown in Fig.~\ref{SchwingerSR}. Our measurement of $I^n_\mathrm{LT}(Q^2)$ 
without the Regge-based parameterization~\cite{Bass:2018uon} for the large-$\nu$ part (open symbols), which is suppressed in $\delta_\mathrm{LT}(Q^2)$, displays a  similar 
pattern as $\delta_\mathrm{LT}^n(Q^2)$. 
The Gerasimov-Drell-Hearn (GDH) relation~\cite{Gerasimov:1965et, Drell:1966jv} can be used to extrapolate our $I^n_\mathrm{LT}(Q^2)$ 
to $Q^2=0$; and provided that the GDH relation is valid, which is widely expected and supported by dedicated experimental
studies~\cite{Helbing:2006zp}, our data satisfy Schwinger's prediction that $I^n_\mathrm{LT}(0)=0$~\cite{Schwinger:1975ti}. 
Our trend contrasts with the MAID model and presumably the $\chi$EFT calculations, since MAID tracks those (see Fig.~\ref{Fig:delta_LT+gamma0}). 
This suggests that the problem lies in the theoretical description of the neutron structure.
The measured $I_\mathrm{LT}(Q^2)$ displays a similar $Q^2$-behavior as $\delta_\mathrm{LT}$, irrespective of the different $\nu$-weighting.
Other integrals without $\nu^{-2}$ weighting formed using our data and reported in~\cite{Sulkosky:2019zmn} 
did not display the surprisingly strong disagreement with the predictions seen here.
 The values of $\gamma_0^n$, $\delta_\mathrm{LT}^n$ and $I_\mathrm{LT}^n$
with their uncertainties are available in the Supplementary Data Files.

Our data indicate that both the TT and LT interferences of the electromagnetic field's components  induce a 
clear spin precession of the neutron. While it was predicted by all calculations and models that  the LT term
influence should intensify at  small $Q^2$, our data reveal the opposite trend.
This notable disagreement is perplexing since our measurements were done
well into the domain where $\chi$EFT is expected to describe reliably
the nucleon properties, especially the ``gold-plated''  $\delta_\mathrm{LT}$.
Lattice QCD calculations of $\delta_\mathrm{LT}(Q^2)$ are possible~\cite{Chambers:2017dov}, but not yet available. 
Our data motivate such calculations since the measured generalized spin-polarizabilities underline a current lack of reliable quantitative descriptions of the Strong Interaction at the nucleon-size scale.

\bigskip\noindent
{\bf{Data availability}}
 All experimental data that support the findings of this study are provided in the Supplementary Data Files or are available from 
 J.P. Chen (jpchen@jlab.org), 
A. Deur (deurpam@jlab.org),
C. Peng (cpeng@jlab.org) or
V. Sulkosky (vasulk@jlab.org)  upon request.

\bigskip\noindent
{\bf{Code availability}}
The computer codes that support the plots within this paper and the findings
of this study are available from 
J.P. Chen (jpchen@jlab.org), 
A. Deur (deurpam@jlab.org),
C. Peng (cpeng@jlab.org) or
V. Sulkosky (vasulk@jlab.org)  upon request.

\bigskip\noindent
{\bf{Author contributions}}
The members of the Jefferson Lab E97-110 Collaboration constructed and operated the experimental
equipment used in this experiment. All authors contributed to the data collection, experiment design and commissioning, data processing, data analysis or Monte Carlo simulations. The following authors especially contributed to the main data analysis: J.P Chen,  A. Deur, C. Peng and V. Sulkosky.

\bigskip\noindent
{\bf{Competing interests}}
The authors declare no competing interests.

\begin{acknowledgments}
We acknowledge the outstanding support of the Jefferson Lab Hall A technical staff  and the Physics and Accelerator 
Divisions that made this work possible. 
We thank A. Deltuva, J. Golak, F. Hagelstein, H. Krebs, V. Lensky, U.-G. Mei{\ss}ner, V. Pascalutsa, 
G. Salm\`e, S. Scopetta and M. Vanderhaeghen for useful discussions and for sharing their calculations. 
We are grateful to V. Pascalutsa and M. Vanderhaeghen for suggesting to compare the data to the Schwinger relation.  
This material is based  upon work supported by the U.S. Department of Energy, Office of Science, 
Office of Nuclear Physics under contract DE-AC05-06OR23177, and by the NSF under grant PHY-0099557.
\end{acknowledgments}

\bibliography{scibib}

\bibliographystyle{Science}

\newpage

\flushleft {\underline{\large{\bf{Data tables}}}}
\begin{table*}[!h]
\begin{center}
{\footnotesize{
\begin{tabular}{|c|c|c|c|c|} \hline
$Q^2$ & $\nu_{max}$ ($W_{max}$)   &  $\delta_\mathrm{LT}^{^3He,~res}(Q^2)\pm$(stat)$\pm$(syst)  & $\delta_\mathrm{LT}^{n,~res}(Q^2)\pm$(stat)$\pm$(syst)   & $\delta_\mathrm{LT}^n(Q^2)\pm$(stat)$\pm$(syst) \\ 
$[$GeV$^2]$ & [GeV]&  [$10^{-4}$ fm$^4$] & [$10^{-4}$ fm$^4$] & [$10^{-4}$ fm$^4$] \\ \hline 
0.035                       & 1.690 (2.00)    & $ -0.356 \pm 0.280 \pm 0.583 $ & $ -0.379 \pm 0.326 \pm 0.677 $ & $ -0.383\pm0.326 \pm0.677$  \\
0.057                       &  1.700 (2.00)  &  $ 0.174 \pm 0.061 \pm 0.169 $ & $ 0.229 \pm 0.071 \pm 0.197 $ & $ 0.225\pm0.071 \pm0.197$  \\
0.079                       &  1.710 (2.00) &  $ 0.360 \pm 0.084 \pm 0.168 $ & $ 0.439 \pm 0.098 \pm 0.195 $ &$ 0.435\pm0.098 \pm0.195$ \\
0.100                       &  2.885 (2.49)  &  $ 0.410 \pm 0.072 \pm 0.180 $ & $ 0.493 \pm 0.083 \pm 0.209 $ & $ 0.491\pm0.083 \pm0.209$ \\
0.150                       &   2.910 (2.48) &  $ 0.178 \pm 0.045 \pm 0.149 $ & $ 0.216 \pm 0.053 \pm 0.173 $ & $ 0.215\pm0.053 \pm0.173$ \\
0.200                       &  2.655 (2.38) &   $ 0.091 \pm 0.024 \pm 0.078 $ & $ 0.112 \pm 0.028 \pm 0.091 $ &$ 0.111\pm0.028 \pm0.091$ \\
0.240                       & 2.320 (2.23)   & $ 0.090 \pm 0.017 \pm 0.041 $ & $ 0.110 \pm 0.020 \pm 0.043 $ & $ 0.108\pm0.020 \pm0.043$ \\
 \hline 
\end{tabular}
}}
\caption{ Data and kinematics for $\delta_\mathrm{LT}(Q^2)$. From left to right: 
Four-momentum transfer squared; 
Maximum $\nu$ value experimentally covered (equivalent maximum invariant $W$ ($W = (M^2 + 2M\nu - Q^2)^{\sfrac{1}{2}}$;
$\delta_\mathrm{LT}^{^3He,~res}(Q^2)$ 
measured over the $\nu$ ($W$) range from pion threshold up to maximum $\nu$ ($W$) covered experimentally (mostly the nucleon resonance region)
and before applying the nuclear corrections to extract the neutron information. 
(stat) denotes the statistical uncertainty and (syst) the systematic uncertainty;
Extracted neutron $\delta_\mathrm{LT}^{n,~res}$ (resonance) after applying nuclear corrections to the previous column;
Total $\delta_\mathrm{LT}^n(Q^2)$.
Comparing the two last columns shows that the unmeasured  parts of $\delta_\mathrm{LT}^n(Q^2)$, i.e, the contributions for $\nu > \nu_{max}$, are negligible.}
\label{tab:delta_LT} 
\end{center} 
\end{table*}

\begin{table*}
\begin{center}
{\footnotesize{
\begin{tabular}{|c|c|c|c|c|} \hline
$Q^2$ &  $\nu_{max}$ ($W_{max}$)  &  $\gamma_0^{^3He,~res}(Q^2)\pm$(stat)$\pm$(syst)  & $\gamma_0^{n,~res}(Q^2)\pm$(stat)$\pm$(syst)   & $\gamma_0^n(Q^2)\pm$(stat)$\pm$(syst)  \\ 
$[$GeV$^2]$ & [GeV]&  [$10^{-4}$ fm$^4$] & [$10^{-4}$ fm$^4$] & [$10^{-4}$ fm$^4$] \\ \hline 
0.035                       &  1.690 (2.00)  & $ -2.590 \pm 0.111 \pm 0.225 $ & $  -3.092 \pm0.129 \pm0.270$ & $-3.094 \pm0.129 \pm0.270$ \\
0.057                       &   1.700 (2.00) &  $ -2.613 \pm 0.121 \pm 0.215 $ & $ -3.115  \pm0.141 \pm0.259 $ &  $-3.117 \pm0.141 \pm0.259$ \\
0.079                       &   1.710 (2.00) &  $  -2.274\pm 0.121 \pm 0.226 $ & $ -2.715  \pm0.140 \pm0.270$ &   $-2.717 \pm0.140 \pm0.270$ \\
0.100                       &  2.885 (2.49)  &  $  -1.725\pm 0.063 \pm 0.143 $ & $ -2.070  \pm0.074 \pm0.170$ &  $-2.070 \pm0.074 \pm0.170$ \\
0.150                       &  2.910 (2.48)  &  $  -1.135\pm 0.044 \pm 0.105 $ & $  -1.370 \pm0.051 \pm0.125$ & $-1.370 \pm0.051 \pm0.125$ \\
0.200                       &  2.655 (2.38) &   $  -0.798 \pm 0.027 \pm 0.056 $ & $ -0.964  \pm0.032 \pm0.065$ &  $-0.965 \pm0.032 \pm0.065$ \\
0.240                       &  2.320 (2.23)  &  $  -0.612 \pm 0.022 \pm 0.043 $ & $ -0.740  \pm0.026 \pm0.050$ &  $-0.742 \pm0.026 \pm0.050$ \\
 \hline 
\end{tabular}
}}
\caption{ Data and kinematics for $\gamma_0(Q^2)$. From left to right: 
Four-momentum transfer squared; 
Maximum $\nu$ value experimentally covered (equivalent maximum invariant $W$ ($W = (M^2 + 2M\nu - Q^2)^{\sfrac{1}{2}}$;
$\gamma_0^{^3He,~res}(Q^2)$
measured over the $\nu$ ($W$) range from pion threshold up to maximum $\nu$ ($W$) covered experimentally (mostly the nucleon resonance region)
and before applying the nuclear corrections to extract the neutron information.
(stat) denotes the statistical uncertainty and (syst) the systematic uncertainty;
Extracted neutron $\gamma_0^{n,~res}$ (resonance) after applying nuclear corrections to the previous column;
Total $\gamma_0^n(Q^2)$;
Comparing the two last columns shows that the unmeasured  parts of $\gamma_0^n(Q^2)$, i.e, the contributions for $\nu > \nu_{max}$, are negligible.
}
\label{tab:gamma_0} 
\end{center} 
\end{table*}

\begin{table*}
\begin{center}
{\footnotesize{
\begin{tabular}{|c|c|c|c|c|c|} \hline
$Q^2$ &$\nu_{max}$ ($W_{max}$)    &  $I_\mathrm{LT}^{^3He,~res}(Q^2)\pm$(stat)$\pm$(syst)  &  $I_\mathrm{LT}^{n,~res}(Q^2)\pm$(stat)$\pm$(syst)  & $I_\mathrm{LT}(Q^2)^n\pm$(stat)$\pm$(syst)  \\ 
$[$GeV$^2]$ & [GeV]& & & \\ \hline 
0.035                       & 1.690 (2.00)    & $-0.326 \pm 0.272 \pm0.520 $ & $-0.294 \pm 0.316 \pm0.604 $  &$-1.112\pm0.316 \pm0.606$ \\
0.057                       & 1.700 (2.00)   &  $0.285 \pm 0.117 \pm0.333 $ & $0.413 \pm 0.136 \pm0.387 $  & $-0.862\pm0.136 \pm0.389$  \\ 
0.079                       & 1.710 (2.00)   &  $0.610 \pm 0.128 \pm0.268 $ & $0.786 \pm 0.149 \pm0.312 $  & $-0.721\pm0.149 \pm0.314$  \\
0.100                       & 2.885 (2.49)   &  $0.879 \pm 0.098 \pm0.280 $ & $1.092 \pm 0.114 \pm0.327 $ & $-0.126\pm0.114 \pm0.329$  \\
0.150                       & 2.910 (2.48)   &  $0.339 \pm 0.049 \pm0.198 $ & $0.458 \pm 0.057 \pm0.231 $ &  $-0.266\pm0.057 \pm0.233$  \\
0.200                       & 2.655 (2.38)  &   $0.289 \pm 0.047 \pm0.228 $ & $0.389 \pm 0.055 \pm0.265 $ &  $-0.345\pm0.055 \pm0.267$  \\
0.240                       & 2.320 (2.23)   &  $0.391 \pm 0.057 \pm0.163 $ & $0.511 \pm 0.067 \pm0.190 $ &  $-0.267\pm0.067 \pm0.192$  \\
 \hline 
\end{tabular}
}}
\caption{ Data and kinematics for $I_\mathrm{LT}(Q^2)$. From left to right: 
Four-momentum transfer squared; 
Maximum $\nu$ value experimentally covered (equivalent maximum invariant $W$ ($W = (M^2 + 2M\nu - Q^2)^{\sfrac{1}{2}}$;
$I_\mathrm{LT}^{^3He,~res}(Q^2)$ 
measured over the $\nu$ ($W$) range from pion threshold up to maximum $\nu$ ($W$) covered experimentally (mostly the nucleon resonance region)
and before applying the nuclear corrections to extract the neutron information.
(stat) denotes the statistical uncertainty and (syst) the systematic uncertainty;
Extracted neutron $I_\mathrm{LT}^{n,~res}$ (resonance) after applying nuclear corrections to the previous column;
Total $I_\mathrm{LT}^n(Q^2)$, including an estimate for the unmeasured contribution above $ \nu_{max}$.
}
\label{tab:I_LT}
\end{center}
\end{table*}

\begin{table*}
\begin{center}
{\footnotesize{
\begin{tabular}{|c|c|c|c|c|c|c|c|c|c|c|c|} \hline
$Q^2$ & $\nu$ & $W$  & $x$  & $\sigma_\mathrm{LT}$ & Stat. & Uncor. syst. & Cor. syst. & $\sigma_\mathrm{TT}$ & Stat. & Uncor. syst. & Cor. syst.\\ 
$[$GeV$^2]$ & [MeV]& [MeV]& &  [$\mu$b] & [$\mu$b] & [$\mu$b] & [$\mu$b]  & [$\mu$b]   & [$\mu$b]   & [$\mu$b]  &  [$\mu$b]\\ \hline 
0.035 &  167.5 &  1076.9 &  0.1114 &  -29.593 &  7.821 &  4.430 &  6.567 &  -50.779 &  8.849 &  4.803 &  7.200 \\
0.035 &  210.0 &  1113.3 &  0.0888 &   -9.283 &  8.641 &  4.463 &  2.718 &  -60.029 &  9.257 &  4.828 &  3.067 \\
0.035 &  250.0 &  1146.5 &  0.0746 &    5.012 &  7.507 &  7.804 &  4.201 &  -70.097 &  7.608 & 10.592 &  5.505 \\
0.035 &  290.0 &  1178.8 &  0.0643 &   -4.189 &  8.133 &  4.768 &  4.598 & -144.587 &  7.398 &  6.008 &  5.937 \\
0.035 &  330.0 &  1210.2 &  0.0565 &  -11.970 &  7.757 &  5.264 &  4.145 & -180.197 &  6.794 &  8.238 &  5.752 \\
0.035 &  370.0 &  1240.8 &  0.0504 &    3.328 &  7.025 &  5.624 &  3.348 & -159.284 &  6.028 &  7.449 &  5.188 \\
0.035 &  410.0 &  1270.7 &  0.0455 &    7.802 &  6.144 &  4.136 &  2.339 & -102.038 &  5.192 &  5.252 &  3.146 \\
0.035 &  450.0 &  1299.9 &  0.0414 &    2.569 &  6.034 &  1.754 &  1.333 &  -56.856 &  4.689 &  3.063 &  1.734 \\
0.035 &  490.0 &  1328.5 &  0.0381 &   -2.776 &  6.006 &  1.077 &  1.003 &  -36.599 &  4.210 &  2.488 &  1.797 \\
0.035 &  530.0 &  1356.4 &  0.0352 &   -0.849 &  5.748 &  0.699 &  2.120 &  -22.279 &  3.827 &  1.881 &  2.340 \\
0.035 &  570.0 &  1383.8 &  0.0327 &    4.796 &  6.483 &  0.646 &  0.845 &  -17.020 &  4.317 &  1.184 &  2.026 \\
0.035 &  610.0 &  1410.7 &  0.0306 &    7.627 &  9.229 &  0.833 &  0.899 &  -19.965 &  7.244 &  1.079 &  2.128 \\
0.035 &  650.0 &  1437.0 &  0.0287 &   11.711 & 10.295 &  0.878 &  0.757 &  -13.732 &  8.808 &  0.690 &  1.336 \\
0.035 &  690.0 &  1462.9 &  0.0270 &   22.036 & 10.449 &  1.301 &  1.731 &  -18.433 &  8.192 &  0.832 &  1.271 \\
0.035 &  730.0 &  1488.4 &  0.0256 &   10.433 & 10.676 &  1.255 &  3.585 &  -17.674 &  6.646 &  0.939 &  1.543 \\
0.035 &  770.0 &  1513.4 &  0.0242 &  -16.164 &  9.879 &  0.690 &  2.002 &   -9.317 &  5.897 &  0.830 &  1.388 \\
0.035 &  810.0 &  1538.0 &  0.0230 &   -3.384 & 10.144 &  0.905 &  2.704 &  -13.691 &  5.873 &  1.137 &  1.231 \\
0.035 &  850.0 &  1562.2 &  0.0219 &    6.799 & 11.135 &  0.939 &  3.434 &  -16.569 &  6.153 &  1.448 &  2.080 \\
0.035 &  890.0 &  1586.0 &  0.0210 &   -5.700 & 10.532 &  0.600 &  3.009 &   -9.035 &  5.791 &  0.515 &  2.530 \\
0.035 &  930.0 &  1609.5 &  0.0201 &   17.363 & 10.209 &  0.852 &  1.794 &  -18.818 &  5.628 &  0.690 &  1.982 \\
0.035 &  970.0 &  1632.7 &  0.0192 &    8.494 &  9.638 &  0.846 &  1.477 &  -21.969 &  4.806 &  1.291 &  0.844 \\
0.035 & 1010.0 &  1655.5 &  0.0185 &   -8.975 &  8.213 &  0.660 &  1.582 &  -25.754 &  4.428 &  1.639 &  1.197 \\
0.035 & 1050.0 &  1678.0 &  0.0178 &  -14.309 &  7.307 &  0.924 &  1.935 &  -15.582 &  4.476 &  1.424 &  1.432 \\
0.035 & 1090.0 &  1700.2 &  0.0171 &    3.170 &  8.109 &  0.537 &  2.119 &  -13.869 &  4.743 &  0.862 &  0.833 \\
0.035 & 1130.0 &  1722.2 &  0.0165 &   -8.033 &  9.470 &  0.818 &  2.159 &   -8.614 &  5.365 &  0.660 &  0.765 \\
0.035 & 1170.0 &  1743.8 &  0.0159 &  -15.669 &  9.088 &  0.909 &  2.006 &  -14.402 &  5.160 &  2.370 &  1.013 \\
 \end{tabular}
}}
\end{center}
\end{table*}
\pagebreak
\begin{table*}
\begin{center}
{\footnotesize{
\begin{tabular}{|c|c|c|c|c|c|c|c|c|c|c|c|}
0.035 & 1210.0 &  1765.2 &  0.0154 &   -3.340 &  9.120 & ~~ 0.257 ~~ &  ~~~  0.488  ~~~ &  -20.220 &  5.268 &   ~~~~   1.405   ~~~~  & ~ 2.551 ~ \\
0.035 & 1250.0 &  1786.3 &  0.0149 &    8.626 &  9.670 &  0.657 &  1.584 &  -17.874 &  5.217 &  0.934 &  1.967 \\
0.035 & 1290.0 &  1807.2 &  0.0145 &    6.658 &  9.509 &  0.504 &  0.796 &  -22.003 &  5.059 &  1.209 &  1.012 \\
0.035 & 1330.0 &  1827.9 &  0.0140 &   10.562 &  9.728 &  0.755 &  1.773 &  -17.263 &  5.284 &  0.894 &  0.558 \\
0.035 & 1370.0 &  1848.3 &  0.0136 &   -6.227 & 10.036 &  0.190 &  1.003 &  -15.983 &  5.941 &  0.877 &  0.438 \\
0.035 & 1410.0 &  1868.5 &  0.0132 &   -0.496 & 10.253 &  0.396 &  1.494 &  -17.223 &  6.110 &  0.925 &  0.429 \\
0.035 & 1450.0 &  1888.5 &  0.0129 &    9.730 & 10.555 &  0.963 &  2.563 &  -17.723 &  6.335 &  0.921 &  0.561 \\
0.035 & 1490.0 &  1908.2 &  0.0125 &  -16.688 & 10.365 &  0.758 &  1.789 &  -15.680 &  6.315 &  0.938 &  0.887 \\
0.035 & 1530.0 &  1927.8 &  0.0122 &   -6.735 & 10.420 &  0.298 &  0.434 &  -26.546 &  6.354 &  1.474 &  1.010 \\
0.035 & 1570.0 &  1947.2 &  0.0119 &   -3.131 & 11.567 &  0.146 &  0.590 &  -16.390 &  6.927 &  0.895 &  0.547 \\
0.035 & 1610.0 &  1966.4 &  0.0116 &  -12.388 & 12.041 &  0.576 &  1.025 &  -15.621 &  7.166 &  0.876 &  0.443 \\
0.035 & 1650.0 &  1985.4 &  0.0113 &  -25.057 & 12.833 &  1.492 &  2.250 &  -11.943 &  7.797 &  0.939 &  0.488 \\
0.035 & 1690.0 &  2004.2 &  0.0110 &   12.734 & 13.111 &  0.708 &  1.519 &  -16.559 &  8.754 &  2.147 &  0.546 \\
0.035 & 1730.0 &  2022.8 &  0.0108 &   -3.404 & 12.912 &  0.180 &  0.339 &  -23.472 &  8.469 &  1.405 &  0.496 \\
0.035 & 1770.0 &  2041.3 &  0.0105 &  -18.275 & 10.605 &  0.945 &  0.752 &  -10.813 &  8.329 &  0.977 &  0.719 \\
0.035 & 1810.0 &  2059.6 &  0.0103 &   -6.756 & 11.453 &  0.312 &  1.048 &  -23.516 &  7.914 &  1.303 &  1.315 \\
0.035 & 1850.0 &  2077.7 &  0.0101 &  -18.357 & 14.142 &  1.001 &  0.658 &   -1.761 & 10.219 &  0.197 &  0.559 \\
0.057 &  177.5 &  1075.4 &  0.1712 &    7.220 & 15.736 &  3.537 &  8.714 &  -92.373 & 19.571 &  4.724 & 11.368 \\
0.057 &  220.0 &  1111.8 &  0.1381 &    8.333 &  7.607 &  3.298 &  3.970 &  -85.011 &  9.007 &  3.795 &  4.741 \\
0.057 &  260.0 &  1145.1 &  0.1168 &    4.145 &  6.307 &  3.452 &  5.860 &  -85.636 &  6.975 &  3.884 &  6.973 \\
0.057 &  300.0 &  1177.4 &  0.1013 &  -11.320 &  6.279 &  4.864 &  6.505 & -136.987 &  6.600 &  5.833 &  7.964 \\
0.057 &  340.0 &  1208.9 &  0.0893 &    3.996 &  6.344 &  5.770 &  5.386 & -179.319 &  6.447 &  7.269 &  6.118 \\
0.057 &  380.0 &  1239.5 &  0.0799 &   16.750 &  5.973 &  6.692 &  6.489 & -159.668 &  5.916 &  7.136 &  6.346 \\
0.057 &  420.0 &  1269.5 &  0.0723 &   14.794 &  5.623 &  4.770 &  4.507 & -104.990 &  5.277 &  4.722 &  3.989 \\
0.057 &  460.0 &  1298.7 &  0.0660 &   16.454 &  5.864 &  2.330 &  2.421 &  -64.416 &  5.024 &  3.446 &  3.120 \\
0.057 &  500.0 &  1327.3 &  0.0608 &   13.597 &  5.883 &  1.659 &  4.063 &  -36.795 &  4.674 &  2.493 &  2.566 \\
 \end{tabular}
}}
\end{center}
\end{table*}
\pagebreak
\begin{table*}
\begin{center}
{\footnotesize{
\begin{tabular}{|c|c|c|c|c|c|c|c|c|c|c|c|}
0.057 &  540.0 &  1355.2 &  0.0562 &    4.128 &  5.647 &  ~~~ 0.697 ~~~  & ~~~  2.973  ~~~ &  -15.701 &  4.370 & ~~~~  0.971 ~~~~ & ~ 2.064 ~ \\
0.057 &  580.0 &  1382.7 &  0.0524 &   -3.473 &  5.804 &  0.581 &  4.200 &  -11.937 &  4.380 &  0.789 &  2.328 \\
0.057 &  620.0 &  1409.5 &  0.0490 &   -3.249 &  6.656 &  0.632 &  1.316 &  -20.045 &  4.979 &  1.075 &  1.765 \\
0.057 &  660.0 &  1435.9 &  0.0460 &    2.172 &  7.360 &  0.474 &  2.020 &  -12.806 &  5.690 &  0.642 &  2.699 \\
0.057 &  700.0 &  1461.8 &  0.0434 &   13.961 &  7.318 &  0.966 &  1.970 &  -11.075 &  5.491 &  0.483 &  2.133 \\
0.057 &  740.0 &  1487.3 &  0.0410 &   14.570 &  7.427 &  1.118 &  2.889 &  -20.931 &  5.619 &  0.900 &  1.405 \\
0.057 &  780.0 &  1512.3 &  0.0389 &    3.221 &  7.617 &  0.675 &  2.554 &  -16.318 &  5.599 &  0.791 &  1.357 \\
0.057 &  820.0 &  1536.9 &  0.0370 &    5.563 &  7.585 &  0.637 &  1.174 &  -14.062 &  5.286 &  0.659 &  0.750 \\
0.057 &  860.0 &  1561.2 &  0.0353 &    0.817 &  7.583 &  0.307 &  0.947 &  -12.846 &  5.180 &  0.696 &  0.753 \\
0.057 &  900.0 &  1585.0 &  0.0338 &    4.921 &  7.582 &  0.588 &  1.167 &  -17.092 &  5.121 &  0.806 &  1.163 \\
0.057 &  940.0 &  1608.5 &  0.0323 &   16.944 &  8.508 &  0.796 &  1.911 &  -22.979 &  5.428 &  1.003 &  1.584 \\
0.057 &  980.0 &  1631.7 &  0.0310 &   11.996 &  9.503 &  1.179 &  2.745 &  -15.319 &  5.418 &  1.004 &  1.096 \\
0.057 & 1020.0 &  1654.5 &  0.0298 &  -23.536 &  9.559 &  1.079 &  1.610 &   -9.656 &  5.365 &  1.281 &  0.666 \\
0.057 & 1060.0 &  1677.0 &  0.0287 &  -26.072 &  9.871 &  1.414 &  3.111 &   -5.177 &  5.518 &  0.649 &  0.900 \\
0.057 & 1100.0 &  1699.3 &  0.0276 &   -0.363 & 10.045 &  0.119 &  1.423 &   -8.255 &  5.946 &  0.430 &  0.809 \\
0.057 & 1140.0 &  1721.2 &  0.0266 &   -1.114 & 10.216 &  0.145 &  0.423 &  -13.182 &  6.421 &  0.699 &  0.462 \\
0.057 & 1180.0 &  1742.9 &  0.0257 &    0.422 & 10.433 &  0.162 &  0.878 &  -13.824 &  6.438 &  0.720 &  0.435 \\
0.057 & 1220.0 &  1764.3 &  0.0249 &    3.185 & 10.857 &  0.291 &  0.661 &  -13.149 &  6.460 &  0.667 &  0.883 \\
0.057 & 1260.0 &  1785.4 &  0.0241 &   -0.032 & 11.082 &  0.142 &  0.533 &   -9.820 &  6.050 &  0.562 &  0.857 \\
0.057 & 1300.0 &  1806.3 &  0.0234 &    3.530 & 10.636 &  0.393 &  0.546 &  -19.872 &  5.708 &  1.107 &  1.144 \\
0.057 & 1340.0 &  1827.0 &  0.0227 &    7.619 & 10.962 &  0.596 &  1.217 &  -16.451 &  5.826 &  0.862 &  1.034 \\
0.057 & 1380.0 &  1847.4 &  0.0220 &   -3.622 & 11.395 &  0.246 &  0.806 &  -15.794 &  6.415 &  0.868 &  1.215 \\
0.057 & 1420.0 &  1867.6 &  0.0214 &   -0.549 & 11.656 &  0.357 &  1.057 &  -16.219 &  6.442 &  0.876 &  0.915 \\
0.057 & 1460.0 &  1887.6 &  0.0208 &    7.762 & 11.952 &  0.710 &  1.369 &  -16.017 &  6.482 &  0.825 &  0.515 \\
0.057 & 1500.0 &  1907.4 &  0.0203 &  -10.832 & 11.640 &  0.617 &  0.986 &  -13.847 &  6.253 &  0.836 &  0.717 \\
0.057 & 1540.0 &  1927.0 &  0.0197 &   -5.680 & 11.515 &  0.254 &  0.371 &  -21.179 &  6.064 &  1.181 &  0.650 \\
 \end{tabular}
}}
\end{center}
\end{table*}
\pagebreak
\begin{table*}
\begin{center}
{\footnotesize{
\begin{tabular}{|c|c|c|c|c|c|c|c|c|c|c|c|}
0.057 & 1580.0 &  1946.3 &  0.0192 &   -1.858 & 12.504 &  ~~~ 0.162  ~~~ &  ~~~ 0.468  ~~~ &  -14.862 &  6.311 &  ~~~ 0.812  ~~~ & ~ 0.678 ~\\
0.057 & 1620.0 &  1965.5 &  0.0188 &   -5.926 & 12.726 &  0.461 &  0.789 &  -14.826 &  6.284 &  0.829 &  0.619 \\
0.057 & 1660.0 &  1984.5 &  0.0183 &  -15.284 & 13.152 &  1.151 &  2.330 &  -11.164 &  6.518 &  0.793 &  0.594 \\
0.057 & 1700.0 &  2003.4 &  0.0179 &    5.253 & 13.024 &  0.574 &  1.241 &  -16.632 &  6.962 &  1.697 &  1.296 \\
0.057 & 1740.0 &  2022.0 &  0.0175 &   -2.795 & 12.474 &  0.177 &  0.293 &  -21.069 &  6.512 &  1.227 &  0.648 \\
0.057 & 1780.0 &  2040.5 &  0.0171 &  -11.946 &  9.985 &  0.612 &  1.899 &  -11.633 &  6.119 &  0.893 &  0.878 \\
0.057 & 1820.0 &  2058.8 &  0.0167 &   -4.724 & 10.712 &  0.237 &  0.985 &  -18.421 &  5.839 &  1.025 &  1.079 \\
0.057 & 1860.0 &  2076.9 &  0.0163 &  -11.979 & 13.047 &  0.712 &  2.817 &   -4.903 &  7.290 &  0.348 &  0.752 \\
0.079 &  187.5 &  1073.9 &  0.2246 &   34.143 & 18.241 &  3.088 &  7.411 &  -64.904 & 24.672 &  5.429 & 10.679 \\
0.079 &  230.0 &  1110.4 &  0.1830 &   17.772 & 12.184 &  3.035 &  3.783 &  -92.997 & 15.691 &  4.322 &  5.277 \\
0.079 &  270.0 &  1143.7 &  0.1559 &   -2.744 &  8.866 &  3.869 &  4.971 & -111.705 & 10.617 &  4.587 &  5.878 \\
0.079 &  310.0 &  1176.0 &  0.1358 &  -12.400 &  7.942 &  5.274 &  5.393 & -154.179 &  9.071 &  6.282 &  6.244 \\
0.079 &  350.0 &  1207.5 &  0.1203 &   11.408 &  8.144 &  6.164 &  5.841 & -177.788 &  9.089 &  7.241 &  6.483 \\
0.079 &  390.0 &  1238.2 &  0.1079 &   34.764 &  7.864 &  6.817 &  4.885 & -157.234 &  8.585 &  7.005 &  4.887 \\
0.079 &  430.0 &  1268.2 &  0.0979 &   33.358 &  7.078 &  4.791 &  2.725 & -105.957 &  7.208 &  4.616 &  2.363 \\
0.079 &  470.0 &  1297.4 &  0.0896 &   22.314 &  6.960 &  2.869 &  1.047 &  -66.394 &  6.410 &  3.677 &  1.380 \\
0.079 &  510.0 &  1326.0 &  0.0825 &   18.464 &  6.282 &  1.899 &  1.063 &  -36.957 &  5.451 &  2.386 &  1.736 \\
0.079 &  550.0 &  1354.0 &  0.0765 &   13.599 &  5.490 &  1.138 &  0.957 &  -17.115 &  4.625 &  0.778 &  1.702 \\
0.079 &  590.0 &  1381.5 &  0.0714 &   -5.347 &  6.480 &  0.826 &  1.173 &  -11.003 &  5.262 &  0.910 &  1.139 \\
0.079 &  630.0 &  1408.4 &  0.0668 &   -9.027 &  6.974 &  0.677 &  2.375 &  -17.658 &  5.457 &  1.287 &  1.825 \\
0.079 &  670.0 &  1434.8 &  0.0628 &   -3.163 &  7.671 &  0.457 &  2.089 &  -15.417 &  5.870 &  0.967 &  2.565 \\
0.079 &  710.0 &  1460.7 &  0.0593 &    6.803 &  7.476 &  0.640 &  2.526 &   -7.954 &  5.633 &  0.483 &  2.015 \\
0.079 &  750.0 &  1486.2 &  0.0561 &   14.630 &  7.207 &  1.060 &  2.058 &  -18.739 &  5.307 &  1.055 &  1.122 \\
0.079 &  790.0 &  1511.2 &  0.0533 &    9.856 &  7.011 &  0.892 &  1.908 &  -16.028 &  4.978 &  1.029 &  1.266 \\
0.079 &  830.0 &  1535.9 &  0.0507 &    9.177 &  6.570 &  0.857 &  1.488 &  -13.424 &  4.470 &  1.107 &  0.843 \\
0.079 &  870.0 &  1560.1 &  0.0484 &    0.695 &  6.005 &  0.333 &  1.137 &   -9.367 &  4.057 &  0.889 &  0.881 \\
 \end{tabular}
}}
\end{center}
\end{table*}
\pagebreak
\begin{table*}
\begin{center}
{\footnotesize{
\begin{tabular}{|c|c|c|c|c|c|c|c|c|c|c|c|}
0.079 &  910.0 &  1584.0 &  0.0463 &    5.882 &  5.417 & ~~~  0.564 ~~~ &  ~~~ 0.961 ~~~ &  -12.198 &  3.555 &  ~~~ 1.055  ~~~ & ~ 0.818 ~\\
0.079 &  950.0 &  1607.5 &  0.0443 &   11.564 &  6.316 &  4.244 &  0.788 &  -13.436 &  4.176 &  1.958 &  1.268 \\
0.079 &  990.0 &  1630.7 &  0.0425 &   12.303 & 10.534 &  1.106 &  2.812 &  -12.655 &  5.667 &  0.773 &  1.305 \\
0.079 & 1030.0 &  1653.5 &  0.0409 &  -21.372 & 10.231 &  0.955 &  2.364 &   -6.450 &  5.427 &  1.054 &  1.287 \\
0.079 & 1070.0 &  1676.1 &  0.0393 &  -26.240 & 10.201 &  1.367 &  3.819 &   -3.708 &  5.380 &  0.610 &  1.147 \\
0.079 & 1110.0 &  1698.3 &  0.0379 &   -2.104 & 10.257 &  0.121 &  1.186 &   -7.792 &  5.619 &  0.419 &  1.100 \\
0.079 & 1150.0 &  1720.3 &  0.0366 &    0.965 & 10.455 &  0.200 &  2.524 &  -14.070 &  5.978 &  0.712 &  1.339 \\
0.079 & 1190.0 &  1742.0 &  0.0354 &    2.804 & 10.560 &  0.224 &  3.987 &  -14.248 &  5.945 &  0.702 &  1.186 \\
0.079 & 1230.0 &  1763.4 &  0.0342 &    5.423 & 10.901 &  0.392 &  1.310 &  -13.352 &  5.970 &  0.643 &  2.508 \\
0.079 & 1270.0 &  1784.5 &  0.0331 &    2.133 & 11.226 &  0.318 &  1.516 &  -10.841 &  5.678 &  0.639 &  0.856 \\
0.079 & 1310.0 &  1805.4 &  0.0321 &    1.789 &  9.723 &  0.412 &  0.904 &  -23.168 &  4.838 &  1.335 &  2.538 \\
0.079 & 1350.0 &  1826.1 &  0.0312 &    5.894 &  9.835 &  0.557 &  1.305 &  -19.972 &  4.818 &  1.087 &  2.263 \\
0.079 & 1390.0 &  1846.5 &  0.0303 &    0.030 & 10.081 &  0.415 &  1.938 &  -19.555 &  5.131 &  1.049 &  2.583 \\
0.079 & 1430.0 &  1866.8 &  0.0294 &    1.654 & 10.405 &  0.479 &  1.437 &  -19.315 &  5.292 &  1.035 &  1.709 \\
0.079 & 1470.0 &  1886.8 &  0.0286 &    6.673 & 10.523 &  0.579 &  1.348 &  -16.243 &  5.254 &  0.834 &  0.761 \\
0.079 & 1510.0 &  1906.5 &  0.0279 &   -4.686 & 10.171 &  0.596 &  1.309 &  -13.888 &  5.054 &  0.809 &  0.828 \\
0.079 & 1550.0 &  1926.1 &  0.0272 &   -4.896 &  9.978 &  0.225 &  0.786 &  -17.706 &  4.908 &  0.988 &  0.511 \\
0.079 & 1590.0 &  1945.5 &  0.0265 &   -0.354 & 10.333 &  0.202 &  0.455 &  -14.538 &  4.777 &  0.786 &  0.801 \\
0.079 & 1630.0 &  1964.7 &  0.0258 &   -0.897 & 10.428 &  0.442 &  1.642 &  -15.400 &  4.767 &  0.838 &  0.931 \\
0.079 & 1670.0 &  1983.7 &  0.0252 &   -6.199 & 10.704 &  0.993 &  2.901 &  -11.479 &  4.933 &  0.721 &  0.804 \\
0.079 & 1710.0 &  2002.6 &  0.0246 &   -1.247 & 10.500 &  0.496 &  1.681 &  -16.864 &  5.130 &  1.390 &  1.502 \\
0.079 & 1750.0 &  2021.2 &  0.0241 &   -2.692 &  9.746 &  0.187 &  0.367 &  -19.326 &  4.745 &  1.113 &  0.728 \\
0.079 & 1790.0 &  2039.7 &  0.0235 &   -6.890 &  8.195 &  0.376 &  1.963 &  -13.116 &  4.471 &  0.863 &  0.913 \\
0.079 & 1830.0 &  2058.0 &  0.0230 &   -3.057 &  9.057 &  0.200 &  0.908 &  -14.979 &  4.587 &  0.833 &  0.936 \\
0.079 & 1870.0 &  2076.2 &  0.0225 &   -7.990 & 10.082 &  0.521 &  2.570 &   -7.945 &  5.094 &  0.497 &  0.747 \\
0.100 &  202.5 &  1077.2 &  0.2632 &   15.456 &  8.022 &  3.954 &  3.871 &  -76.998 & 11.594 &  6.669 &  7.076 \\
 \end{tabular}
}}
\end{center}
\end{table*}
\pagebreak
\begin{table*}
\begin{center}
{\footnotesize{
\begin{tabular}{|c|c|c|c|c|c|c|c|c|c|c|c|}
0.100 &  245.0 &  1113.6 &  0.2175 &   19.384 &  7.195 &  ~~~ 4.030  ~~~&  ~~~ 3.331  ~~~&  -65.123 &  9.686 &  ~~~ 5.755  ~~~ & ~ 5.130 ~\\
0.100 &  285.0 &  1146.8 &  0.1870 &   26.126 &  5.169 &  3.594 &  3.967 & -101.549 &  6.431 &  4.671 &  5.009 \\
0.100 &  325.0 &  1179.1 &  0.1640 &    9.181 &  5.267 &  5.300 &  4.472 & -145.831 &  6.295 &  6.330 &  5.300 \\
0.100 &  365.0 &  1210.5 &  0.1460 &   15.938 &  5.842 &  8.753 &  4.948 & -185.663 &  6.703 &  9.835 &  5.651 \\
0.100 &  405.0 &  1241.1 &  0.1316 &    9.510 &  5.673 &  5.847 &  4.029 & -138.361 &  6.112 &  6.834 &  4.482 \\
0.100 &  445.0 &  1271.0 &  0.1198 &   21.309 &  5.085 &  3.680 &  1.449 &  -84.572 &  5.254 &  4.195 &  1.821 \\
0.100 &  485.0 &  1300.2 &  0.1099 &   21.822 &  4.835 &  2.374 &  1.142 &  -45.411 &  4.819 &  2.345 &  1.578 \\
0.100 &  525.0 &  1328.7 &  0.1015 &    5.873 &  5.348 &  0.828 &  3.050 &  -17.056 &  5.058 &  0.906 &  2.513 \\
0.100 &  565.0 &  1356.7 &  0.0943 &   11.819 &  4.088 &  1.243 &  2.111 &  -17.313 &  3.736 &  1.101 &  1.720 \\
0.100 &  605.0 &  1384.1 &  0.0881 &   10.628 &  3.602 &  1.331 &  2.131 &  -20.914 &  3.192 &  1.446 &  1.485 \\
0.100 &  645.0 &  1410.9 &  0.0826 &    6.652 &  4.049 &  1.041 &  1.367 &  -22.418 &  3.365 &  1.561 &  1.095 \\
0.100 &  685.0 &  1437.3 &  0.0778 &   10.810 &  4.650 &  1.263 &  0.900 &  -21.760 &  3.758 &  1.737 &  1.250 \\
0.100 &  725.0 &  1463.2 &  0.0735 &    7.994 &  4.797 &  0.884 &  0.550 &  -14.369 &  3.857 &  1.624 &  0.831 \\
0.100 &  765.0 &  1488.6 &  0.0697 &    4.631 &  4.649 &  0.766 &  0.824 &  -13.315 &  3.650 &  1.297 &  0.902 \\
0.100 &  805.0 &  1513.6 &  0.0662 &    8.032 &  4.672 &  1.190 &  0.631 &  -12.766 &  3.494 &  1.563 &  0.877 \\
0.100 &  845.0 &  1538.2 &  0.0631 &    6.553 &  4.473 &  1.124 &  0.330 &  -11.851 &  3.606 &  1.709 &  0.640 \\
0.100 &  885.0 &  1562.4 &  0.0602 &    4.063 &  4.603 &  0.570 &  0.354 &   -7.306 &  3.889 &  1.272 &  0.615 \\
0.100 &  925.0 &  1586.2 &  0.0576 &    3.901 &  4.887 &  0.682 &  0.339 &   -7.039 &  3.701 &  1.351 &  0.521 \\
0.100 &  965.0 &  1609.7 &  0.0552 &    7.498 & 10.407 &  9.152 &  0.687 &   -3.231 &  7.243 &  3.668 &  0.847 \\
0.100 & 1005.0 &  1632.9 &  0.0530 &    8.199 &  6.938 &  0.583 &  0.580 &   -8.522 &  4.879 &  0.443 &  0.958 \\
0.100 & 1045.0 &  1655.7 &  0.0510 &    3.330 &  6.326 &  1.012 &  2.855 &  -19.256 &  4.404 &  2.984 &  1.133 \\
0.100 & 1085.0 &  1678.2 &  0.0491 &   -5.996 &  6.643 &  1.041 &  3.084 &  -13.311 &  4.487 &  1.578 &  1.626 \\
0.100 & 1125.0 &  1700.4 &  0.0474 &    5.958 &  5.880 &  0.558 &  0.716 &  -14.852 &  3.817 &  0.716 &  0.856 \\
0.100 & 1165.0 &  1722.4 &  0.0457 &    5.983 &  6.420 &  0.522 &  3.161 &  -21.348 &  3.982 &  1.217 &  1.772 \\
0.100 & 1205.0 &  1744.0 &  0.0442 &   -1.774 &  6.282 &  0.583 &  2.308 &  -19.196 &  3.834 &  2.291 &  1.536 \\
0.100 & 1245.0 &  1765.4 &  0.0428 &    0.367 &  6.886 &  0.344 &  0.767 &  -16.672 &  4.099 &  1.209 &  2.685 \\ 
\end{tabular}
}}
\end{center}
\end{table*}
\pagebreak
\begin{table*}
\begin{center}
{\footnotesize{
\begin{tabular}{|c|c|c|c|c|c|c|c|c|c|c|c|}
0.100 & 1285.0 &  1786.5 &  0.0415 &    8.289 & ~ 6.606 ~ & ~~~   0.901 ~~~  &  ~~~  2.520  ~~~ & ~ -28.115 ~ &  3.851 &  ~~~ 1.665  ~~~&  ~ 1.862 ~\\
0.100 & 1325.0 &  1807.4 &  0.0402 &   -2.222 &  9.184 &  0.268 &  3.484 &  -10.811 &  4.707 &  0.642 &  1.507 \\
0.100 & 1365.0 &  1828.1 &  0.0390 &    1.185 &  9.278 &  0.396 &  1.450 &   -9.927 &  4.728 &  0.561 &  0.632 \\
0.100 & 1405.0 &  1848.5 &  0.0379 &   -3.194 &  9.330 &  0.228 &  0.502 &  -12.630 &  4.945 &  0.778 &  0.957 \\
0.100 & 1445.0 &  1868.7 &  0.0369 &   -3.062 &  9.625 &  0.281 &  1.343 &   -4.770 &  4.941 &  0.309 &  3.315 \\
0.100 & 1485.0 &  1888.7 &  0.0359 &    1.817 &  9.708 &  0.369 &  1.398 &  -10.809 &  4.950 &  0.625 &  0.566 \\
0.100 & 1525.0 &  1908.4 &  0.0349 &  -14.830 &  9.162 &  0.684 &  2.861 &  -19.837 &  4.816 &  1.353 &  3.762 \\
0.100 & 1565.0 &  1928.0 &  0.0341 &   -2.055 &  9.173 &  0.206 &  0.802 &  -12.951 &  4.850 &  0.779 &  0.856 \\
0.100 & 1605.0 &  1947.4 &  0.0332 &   -4.243 &  9.530 &  0.193 &  1.135 &   -9.340 &  4.905 &  0.600 &  0.611 \\
0.100 & 1645.0 &  1966.5 &  0.0324 &    2.988 &  9.658 &  0.687 &  1.475 &  -17.478 &  5.030 &  1.073 &  1.227 \\
0.100 & 1685.0 &  1985.5 &  0.0316 &    0.209 &  7.303 &  0.857 &  1.807 &  -12.793 &  3.697 &  0.748 &  0.948 \\
0.100 & 1725.0 &  2004.3 &  0.0309 &   -6.303 &  7.199 &  0.399 &  1.489 &  -17.072 &  3.758 &  1.158 &  0.984 \\
0.100 & 1765.0 &  2023.0 &  0.0302 &   -2.988 &  7.012 &  0.190 &  0.278 &  -17.894 &  3.704 &  1.000 &  0.489 \\
0.100 & 1805.0 &  2041.5 &  0.0295 &   -2.567 &  6.152 &  0.222 &  0.989 &  -14.275 &  3.386 &  0.827 &  0.351 \\
0.100 & 1845.0 &  2059.8 &  0.0289 &   -2.509 &  7.388 &  0.206 &  0.900 &  -12.301 &  3.916 &  0.688 &  0.376 \\
0.100 & 1885.0 &  2077.9 &  0.0283 &   -4.084 &  7.543 &  0.436 &  1.750 &   -9.504 &  3.958 &  0.560 &  0.562 \\
0.100 & 2005.0 &  2131.4 &  0.0266 &    3.842 &  4.597 &  0.274 &  0.536 &  -12.950 &  2.610 &  0.675 &  0.322 \\
0.100 & 2045.0 &  2148.9 &  0.0261 &    3.536 &  4.544 &  0.289 &  0.070 &  -12.664 &  2.563 &  0.671 &  0.119 \\
0.100 & 2085.0 &  2166.3 &  0.0256 &    3.312 &  4.672 &  0.275 &  0.081 &  -12.602 &  2.601 &  0.669 &  0.123 \\
0.100 & 2125.0 &  2183.6 &  0.0251 &    3.729 &  4.758 &  0.291 &  0.203 &  -12.564 &  2.613 &  0.666 &  0.132 \\
0.100 & 2165.0 &  2200.7 &  0.0246 &    3.467 &  6.179 &  0.425 &  1.015 &  -12.002 &  3.365 &  0.643 &  0.216 \\
0.100 & 2205.0 &  2217.7 &  0.0242 &   -7.082 &  9.019 &  0.380 &  0.589 &  -11.031 &  4.931 &  0.664 &  0.175 \\
0.100 & 2245.0 &  2234.6 &  0.0237 &    3.104 &  7.672 &  0.268 &  0.255 &  -15.142 &  4.189 &  0.852 &  0.169 \\
0.100 & 2285.0 &  2251.3 &  0.0233 &    1.188 &  7.367 &  0.201 &  0.225 &  -14.907 &  4.014 &  0.850 &  0.208 \\
0.100 & 2325.0 &  2267.9 &  0.0229 &    0.464 &  4.258 &  0.150 &  0.231 &  -12.881 &  2.418 &  0.662 &  0.364 \\
0.100 & 2365.0 &  2284.4 &  0.0225 &    3.874 &  4.411 &  0.287 &  0.739 &  -10.969 &  2.221 &  0.630 &  0.435 \\
\end{tabular}
}}
\end{center}
\end{table*}
\pagebreak
\begin{table*}
\begin{center}
{\footnotesize{
\begin{tabular}{|c|c|c|c|c|c|c|c|c|c|c|c|}
0.100 & 2405.0 &  2300.8 &  0.0222 &   -3.542 & ~ 4.432 ~ &  ~~~  0.404  ~~~ &  ~~~  1.238  ~~~  &  -11.716 &  2.166 &  ~~~  0.681 ~~~  & ~ 0.430 ~ \\
0.100 & 2445.0 &  2317.0 &  0.0218 &   -2.553 &  4.665 &  0.329 &  1.321 &  -10.259 &  2.300 &  0.588 &  0.452 \\
0.100 & 2485.0 &  2333.2 &  0.0214 &   -0.625 &  5.832 &  0.631 &  1.708 &   -8.202 &  3.259 &  0.638 &  0.364 \\
0.100 & 2525.0 &  2349.2 &  0.0211 &   -4.785 &  5.523 &  0.537 &  1.282 &  -12.208 &  3.150 &  0.733 &  0.456 \\
0.100 & 2565.0 &  2365.1 &  0.0208 &   -5.088 &  5.431 &  0.310 &  1.000 &   -8.565 &  2.863 &  0.495 &  0.953 \\
0.100 & 2605.0 &  2380.9 &  0.0205 &   -4.249 &  6.307 &  0.211 &  0.465 &   -7.268 &  3.132 &  0.448 &  1.061 \\
0.100 & 2645.0 &  2396.6 &  0.0201 &   -1.454 &  6.765 &  0.106 &  0.354 &  -12.333 &  3.318 &  0.730 &  0.448 \\
0.100 & 2685.0 &  2412.2 &  0.0198 &   -3.477 &  7.573 &  0.214 &  0.650 &  -13.003 &  3.659 &  0.759 &  0.300 \\
0.100 & 2725.0 &  2427.7 &  0.0196 &   -8.275 &  8.178 &  0.403 &  0.726 &  -13.838 &  4.016 &  0.831 &  0.316 \\
0.100 & 2765.0 &  2443.2 &  0.0193 &   -5.867 &  7.309 &  0.298 &  0.926 &  -15.363 &  4.422 &  0.968 &  1.116 \\
0.100 & 2805.0 &  2458.5 &  0.0190 &   -2.355 &  7.693 &  0.229 &  0.668 &  -15.399 &  5.141 &  1.072 &  1.665 \\
0.100 & 2845.0 &  2473.7 &  0.0187 &   -5.013 &  6.881 &  0.306 &  0.212 &  -12.836 &  4.092 &  0.743 &  0.682 \\
0.100 & 2885.0 &  2488.8 &  0.0185 &   -4.007 &  7.609 &  0.178 &  1.216 &  -15.019 &  4.365 &  0.880 &  1.561 \\
0.150 &  227.5 &  1075.8 &  0.3514 &    3.537 &  6.103 &  9.555 &  8.491 &  -61.256 & 10.170 & 11.998 & 11.833 \\
0.150 &  270.0 &  1112.2 &  0.2961 &   12.933 &  5.666 &  2.324 &  5.877 &  -50.296 &  8.559 &  3.450 &  9.452 \\
0.150 &  310.0 &  1145.5 &  0.2579 &   25.440 &  4.679 &  2.025 &  7.919 &  -94.374 &  6.515 &  3.322 &  9.936 \\
0.150 &  350.0 &  1177.8 &  0.2284 &   10.614 &  4.508 &  3.682 &  6.815 & -143.747 &  5.921 &  4.929 &  8.771 \\
0.150 &  390.0 &  1209.2 &  0.2050 &    4.516 &  4.762 &  5.869 &  6.254 & -172.890 &  5.986 &  7.373 &  7.870 \\
0.150 &  430.0 &  1239.9 &  0.1859 &    5.435 &  5.039 &  4.319 &  4.480 & -131.759 &  6.205 &  5.353 &  5.498 \\
0.150 &  470.0 &  1269.8 &  0.1701 &    9.906 &  4.742 &  3.185 &  1.466 &  -84.240 &  5.769 &  4.149 &  1.659 \\
0.150 &  510.0 &  1299.0 &  0.1567 &    8.116 &  4.447 &  2.427 &  0.564 &  -49.114 &  5.182 &  3.235 &  0.789 \\
0.150 &  550.0 &  1327.6 &  0.1453 &    7.676 &  3.913 &  1.581 &  0.962 &  -28.701 &  4.284 &  2.042 &  1.501 \\
0.150 &  590.0 &  1355.5 &  0.1355 &    8.485 &  3.482 &  1.581 &  0.894 &  -20.635 &  3.634 &  1.832 &  1.424 \\
0.150 &  630.0 &  1383.0 &  0.1269 &    8.687 &  3.196 &  1.663 &  0.799 &  -18.807 &  3.235 &  2.089 &  1.208 \\
0.150 &  670.0 &  1409.8 &  0.1193 &    8.416 &  3.141 &  1.160 &  1.107 &  -16.606 &  3.148 &  1.882 &  1.194 \\
0.150 &  710.0 &  1436.2 &  0.1126 &    6.033 &  3.297 &  0.852 &  0.856 &  -14.093 &  3.397 &  1.168 &  1.285 \\
\end{tabular}
}}
\end{center}
\end{table*}
\pagebreak
\begin{table*}
\begin{center}
{\footnotesize{
\begin{tabular}{|c|c|c|c|c|c|c|c|c|c|c|c|}
0.150 &  750.0 &  1462.1 &  0.1066 &    7.997 &  3.653 &  ~~~  0.433  ~~~ &  ~~~  0.693  ~~~ &  -10.516 & ~ 3.717 ~ &  ~~~  0.663  ~~~ & ~ 1.011 ~ \\
0.150 &  790.0 &  1487.6 &  0.1012 &    9.738 &  4.130 &  0.866 &  1.523 &   -8.556 &  3.763 &  1.072 &  1.499 \\
0.150 &  830.0 &  1512.6 &  0.0963 &    2.912 &  4.650 &  1.160 &  2.293 &   -3.873 &  3.830 &  0.970 &  2.076 \\
0.150 &  870.0 &  1537.2 &  0.0919 &    3.913 &  4.955 &  1.031 &  1.372 &   -5.819 &  4.151 &  1.259 &  1.293 \\
0.150 &  910.0 &  1561.4 &  0.0878 &    1.937 &  5.032 &  0.322 &  1.756 &   -2.842 &  4.165 &  0.739 &  2.246 \\
0.150 &  950.0 &  1585.3 &  0.0841 &    0.456 &  4.813 &  0.523 &  2.733 &   -6.978 &  3.791 &  0.905 &  1.895 \\
0.150 &  990.0 &  1608.8 &  0.0807 &   14.691 &  8.089 &  1.680 &  1.313 &  -16.938 &  5.876 &  1.031 &  1.181 \\
0.150 & 1030.0 &  1631.9 &  0.0776 &    3.467 &  4.850 &  0.282 &  1.390 &   -6.773 &  3.744 &  0.312 &  0.817 \\
0.150 & 1070.0 &  1654.8 &  0.0747 &    1.026 &  4.520 &  0.336 &  3.396 &  -12.652 &  3.690 &  0.675 &  1.371 \\
0.150 & 1110.0 &  1677.3 &  0.0720 &   -9.527 &  4.579 &  0.439 &  0.637 &   -8.883 &  3.715 &  0.701 &  0.726 \\
0.150 & 1150.0 &  1699.5 &  0.0695 &   -2.496 &  4.546 &  0.339 &  1.135 &  -10.505 &  3.456 &  0.620 &  0.495 \\
0.150 & 1190.0 &  1721.5 &  0.0672 &    2.371 &  4.692 &  0.317 &  2.825 &  -13.350 &  3.472 &  0.673 &  1.156 \\
0.150 & 1230.0 &  1743.1 &  0.0650 &   -1.243 &  4.488 &  0.380 &  1.791 &  -13.722 &  3.419 &  1.102 &  0.724 \\
0.150 & 1270.0 &  1764.5 &  0.0629 &    6.832 &  5.027 &  0.665 &  1.436 &  -16.786 &  3.863 &  0.914 &  1.377 \\
0.150 & 1310.0 &  1785.7 &  0.0610 &   10.278 &  6.246 &  1.002 &  1.441 &  -17.229 &  4.181 &  0.953 &  0.635 \\
0.150 & 1350.0 &  1806.6 &  0.0592 &   -5.703 &  6.405 &  0.230 &  1.623 &   -6.508 &  4.207 &  0.473 &  0.739 \\
0.150 & 1390.0 &  1827.2 &  0.0575 &   -3.262 &  6.414 &  0.275 &  0.385 &   -6.172 &  4.219 &  0.424 &  0.589 \\
0.150 & 1430.0 &  1847.7 &  0.0559 &   -1.099 &  6.157 &  0.284 &  0.697 &  -12.787 &  4.285 &  0.806 &  0.952 \\
0.150 & 1470.0 &  1867.8 &  0.0544 &   -4.573 &  7.312 &  0.282 &  0.403 &   -0.691 &  4.506 &  0.127 &  0.559 \\
0.150 & 1510.0 &  1887.8 &  0.0529 &   -1.400 &  7.457 &  0.157 &  0.385 &   -7.804 &  4.531 &  0.507 &  0.342 \\
0.150 & 1550.0 &  1907.6 &  0.0516 &  -13.139 &  6.321 &  0.736 &  0.198 &  -20.054 &  4.098 &  1.409 &  0.202 \\
0.150 & 1590.0 &  1927.2 &  0.0503 &    2.114 &  6.761 &  0.262 &  0.327 &   -9.427 &  4.236 &  0.565 &  0.139 \\
0.150 & 1630.0 &  1946.6 &  0.0490 &   -2.797 &  6.089 &  0.247 &  0.319 &   -7.940 &  3.750 &  0.528 &  0.207 \\
0.150 & 1670.0 &  1965.8 &  0.0479 &    9.262 &  6.393 &  0.764 &  0.528 &  -17.845 &  3.881 &  1.020 &  0.214 \\
0.150 & 1710.0 &  1984.8 &  0.0467 &    3.543 &  4.445 &  0.773 &  2.983 &  -12.193 &  2.476 &  0.661 &  0.843 \\
0.150 & 1750.0 &  2003.6 &  0.0457 &  -10.549 &  4.368 &  0.358 &  1.964 &  -12.722 &  2.418 &  0.789 &  0.593 \\
\end{tabular}
}}
\end{center}
\end{table*}
\pagebreak
\begin{table*}
\begin{center}
{\footnotesize{
\begin{tabular}{|c|c|c|c|c|c|c|c|c|c|c|c|}
0.150 & 1790.0 &  2022.2 &  0.0447 &   -1.536 &  4.307 &  ~~~  0.243  ~~~ &  ~~~  0.227 ~~~  &  -14.143 & ~ 2.386 ~ &  ~~~  0.781 ~~~  & ~ 0.312 ~ \\
0.150 & 1830.0 &  2040.7 &  0.0437 &    3.487 &  3.931 &  0.384 &  0.864 &  -14.068 &  2.187 &  0.743 &  0.288 \\
0.150 & 1870.0 &  2059.0 &  0.0427 &    1.752 &  4.319 &  0.330 &  0.782 &  -13.281 &  2.276 &  0.725 &  0.529 \\
0.150 & 1910.0 &  2077.2 &  0.0419 &   -1.936 &  4.157 &  0.259 &  1.197 &  -11.872 &  2.187 &  0.679 &  0.343 \\
0.150 & 1950.0 &  2095.1 &  0.0410 &    3.002 &  4.098 &  0.427 &  1.225 &  -10.719 &  2.254 &  0.594 &  0.517 \\
0.150 & 1990.0 &  2113.0 &  0.0402 &   -5.769 &  4.182 &  0.278 &  0.932 &   -8.489 &  2.467 &  0.518 &  0.451 \\
0.150 & 2030.0 &  2130.7 &  0.0394 &    0.327 &  4.132 &  0.156 &  0.364 &  -12.323 &  3.232 &  0.679 &  0.221 \\
0.150 & 2070.0 &  2148.2 &  0.0386 &   -0.294 &  4.412 &  0.145 &  0.286 &  -12.112 &  3.424 &  0.675 &  0.137 \\
0.150 & 2110.0 &  2165.6 &  0.0379 &   -0.836 &  4.747 &  0.136 &  0.261 &  -12.090 &  3.619 &  0.679 &  0.139 \\
0.150 & 2150.0 &  2182.9 &  0.0372 &   -1.796 &  6.129 &  0.163 &  0.305 &  -11.343 &  3.367 &  0.646 &  0.148 \\
0.150 & 2190.0 &  2200.0 &  0.0365 &    7.273 &  6.624 &  0.496 &  0.202 &  -11.858 &  3.544 &  0.630 &  0.145 \\
0.150 & 2230.0 &  2217.0 &  0.0358 &    0.350 &  6.497 &  0.130 &  0.355 &  -10.022 &  3.422 &  0.565 &  0.160 \\
0.150 & 2270.0 &  2233.9 &  0.0352 &   -2.228 &  5.215 &  0.119 &  0.253 &  -11.921 &  3.702 &  0.679 &  0.179 \\
0.150 & 2310.0 &  2250.6 &  0.0346 &   -1.973 &  5.038 &  0.124 &  0.367 &  -12.010 &  3.503 &  0.729 &  0.237 \\
0.150 & 2350.0 &  2267.2 &  0.0340 &   -2.392 &  4.845 &  0.146 &  0.486 &  -11.864 &  3.313 &  0.873 &  0.550 \\
0.150 & 2390.0 &  2283.7 &  0.0334 &    6.938 &  5.780 &  0.460 &  1.499 &   -9.429 &  2.599 &  0.548 &  0.737 \\
0.150 & 2430.0 &  2300.1 &  0.0329 &   -7.159 &  5.875 &  0.716 &  2.450 &   -9.369 &  2.500 &  0.612 &  0.803 \\
0.150 & 2470.0 &  2316.3 &  0.0324 &   -7.662 &  5.669 &  0.419 &  2.665 &   -8.457 &  2.455 &  0.526 &  0.718 \\
0.150 & 2510.0 &  2332.5 &  0.0318 &  -11.704 &  5.834 &  0.413 &  3.805 &   -6.864 &  2.889 &  0.452 &  0.703 \\
0.150 & 2550.0 &  2348.5 &  0.0313 &  -12.251 &  5.803 &  0.497 &  3.442 &   -8.156 &  2.970 &  0.566 &  0.695 \\
0.150 & 2590.0 &  2364.4 &  0.0309 &   -2.773 &  6.027 &  0.516 &  2.832 &   -5.731 &  2.812 &  0.359 &  1.226 \\
0.150 & 2630.0 &  2380.3 &  0.0304 &   -1.879 &  6.743 &  0.276 &  1.265 &   -6.109 &  3.010 &  0.386 &  1.347 \\
0.150 & 2670.0 &  2396.0 &  0.0299 &    0.356 &  6.534 &  0.110 &  0.490 &  -11.304 &  2.891 &  0.655 &  0.434 \\
0.150 & 2710.0 &  2411.6 &  0.0295 &   -0.429 &  6.377 &  0.185 &  0.460 &  -11.109 &  2.753 &  0.643 &  0.483 \\
0.150 & 2750.0 &  2427.1 &  0.0291 &   -4.804 &  7.336 &  0.277 &  4.534 &   -9.701 &  2.939 &  0.590 &  0.729 \\
0.150 & 2790.0 &  2442.5 &  0.0287 &    0.726 &  6.995 &  0.418 &  2.757 &  -12.856 &  3.135 &  0.705 &  2.397 \\
\end{tabular}
}}
\end{center}
\end{table*}
\pagebreak
\begin{table*}
\begin{center}
{\footnotesize{
\begin{tabular}{|c|c|c|c|c|c|c|c|c|c|c|c|}
0.150 & 2830.0 &  2457.8 &  0.0282 &    6.860 &  6.432 &  ~~~  0.509 ~~~  &  ~~~  0.653  ~~~ &  -13.259 & ~ 3.183 ~ &  ~~~  0.774  ~~~ & ~ 2.246 ~ \\
0.150 & 2870.0 &  2473.1 &  0.0279 &    1.415 &  5.449 &  0.366 &  0.752 &  -10.079 &  2.562 &  0.629 &  0.588 \\
0.150 & 2910.0 &  2488.2 &  0.0275 &   -3.242 &  5.784 &  0.087 &  1.329 &   -7.754 &  2.637 &  0.496 &  0.747 \\
0.200 &  252.5 &  1074.3 &  0.4221 &   13.790 &  3.318 & 10.709 &  7.560 &  -38.694 &  6.467 & 14.450 & 11.419 \\
0.200 &  295.0 &  1110.8 &  0.3613 &    9.440 &  3.819 &  1.772 &  3.899 &  -67.255 &  6.461 &  2.727 &  7.448 \\
0.200 &  335.0 &  1144.1 &  0.3181 &    6.040 &  3.367 &  1.457 &  4.517 &  -93.382 &  5.187 &  2.519 &  7.215 \\
0.200 &  375.0 &  1176.5 &  0.2842 &    2.222 &  3.248 &  2.957 &  5.386 & -135.555 &  4.669 &  4.306 &  7.699 \\
0.200 &  415.0 &  1207.9 &  0.2568 &    1.626 &  3.529 &  3.853 &  5.221 & -150.585 &  4.786 &  5.154 &  6.992 \\
0.200 &  455.0 &  1238.6 &  0.2342 &    6.285 &  3.749 &  3.791 &  3.749 & -132.563 &  4.892 &  4.953 &  4.900 \\
0.200 &  495.0 &  1268.6 &  0.2153 &   12.098 &  3.308 &  2.779 &  1.806 &  -83.584 &  4.189 &  3.587 &  2.171 \\
0.200 &  535.0 &  1297.8 &  0.1992 &    7.023 &  4.035 &  1.809 &  1.064 &  -48.229 &  4.965 &  2.323 &  1.226 \\
0.200 &  575.0 &  1326.4 &  0.1854 &    6.935 &  5.100 &  1.208 &  0.692 &  -30.576 &  6.097 &  1.489 &  0.859 \\
0.200 &  615.0 &  1354.4 &  0.1733 &    7.341 &  4.967 &  1.082 &  0.486 &  -24.222 &  5.800 &  1.276 &  0.729 \\
0.200 &  655.0 &  1381.8 &  0.1627 &    4.639 &  4.694 &  0.865 &  1.129 &  -18.166 &  5.372 &  1.073 &  1.064 \\
0.200 &  695.0 &  1408.7 &  0.1534 &    3.931 &  4.665 &  0.465 &  1.293 &   -9.967 &  5.240 &  0.648 &  1.100 \\
0.200 &  735.0 &  1435.1 &  0.1450 &    3.131 &  4.572 &  0.514 &  0.634 &  -13.385 &  5.003 &  0.670 &  0.769 \\
0.200 &  775.0 &  1461.0 &  0.1375 &    8.223 &  4.833 &  0.548 &  0.294 &  -11.448 &  4.895 &  0.505 &  0.480 \\
0.200 &  815.0 &  1486.5 &  0.1308 &    5.739 &  4.748 &  0.399 &  0.370 &   -6.351 &  4.691 &  0.421 &  0.376 \\
0.200 &  855.0 &  1511.6 &  0.1247 &   -6.748 &  4.703 &  0.476 &  0.971 &    0.020 &  4.560 &  0.362 &  0.646 \\
0.200 &  895.0 &  1536.2 &  0.1191 &   -2.512 &  4.395 &  0.310 &  1.484 &   -2.180 &  4.243 &  0.223 &  0.887 \\
0.200 &  935.0 &  1560.4 &  0.1140 &    1.714 &  3.861 &  0.168 &  0.450 &   -5.796 &  3.710 &  0.240 &  0.312 \\
0.200 &  975.0 &  1584.3 &  0.1093 &   -2.916 &  3.869 &  0.225 &  0.645 &   -3.320 &  3.575 &  0.233 &  0.430 \\
0.200 & 1015.0 &  1607.8 &  0.1050 &    0.873 &  3.603 &  0.197 &  0.730 &   -2.289 &  3.234 &  0.138 &  0.436 \\
0.200 & 1055.0 &  1631.0 &  0.1010 &    3.411 &  3.369 &  0.260 &  0.408 &   -6.416 &  2.977 &  0.303 &  0.442 \\
0.200 & 1095.0 &  1653.8 &  0.0973 &    0.917 &  3.160 &  0.313 &  0.612 &   -8.377 &  2.874 &  0.424 &  0.417 \\
0.200 & 1135.0 &  1676.4 &  0.0939 &   -5.114 &  3.164 &  0.227 &  0.571 &   -6.155 &  2.851 &  0.423 &  0.278 \\
\end{tabular}
}}
\end{center}
\end{table*}
\pagebreak
\begin{table*}
\begin{center}
{\footnotesize{
\begin{tabular}{|c|c|c|c|c|c|c|c|c|c|c|c|}
0.200 & 1175.0 &  1698.6 &  0.0907 &   -1.376 &  3.306 &  ~~~  0.289  ~~~ &  ~~~  0.562  ~~~  &   -8.777 & ~ 2.679 ~& ~~~   0.469  ~~~ & ~ 0.309 ~ \\
0.200 & 1215.0 &  1720.6 &  0.0877 &   -0.408 &  3.322 &  0.192 &  0.583 &   -7.995 &  2.632 &  0.402 &  0.504 \\
0.200 & 1255.0 &  1742.2 &  0.0849 &   -2.934 &  3.316 &  0.269 &  0.652 &  -10.711 &  2.603 &  0.594 &  0.281 \\
0.200 & 1295.0 &  1763.7 &  0.0823 &    6.539 &  3.535 &  0.628 &  0.385 &  -15.119 &  2.762 &  0.751 &  0.232 \\
0.200 & 1335.0 &  1784.8 &  0.0798 &    4.969 &  4.095 &  0.587 &  0.663 &  -14.784 &  2.905 &  0.781 &  0.283 \\
0.200 & 1375.0 &  1805.7 &  0.0775 &    8.241 &  4.053 &  0.696 &  0.874 &  -17.368 &  3.007 &  0.870 &  1.131 \\
0.200 & 1415.0 &  1826.4 &  0.0753 &    0.926 &  3.820 &  0.250 &  0.803 &  -10.907 &  2.816 &  0.584 &  1.000 \\
0.200 & 1455.0 &  1846.8 &  0.0733 &    3.559 &  3.600 &  0.372 &  0.670 &  -12.090 &  2.645 &  0.596 &  0.728 \\
0.200 & 1495.0 &  1867.0 &  0.0713 &    0.530 &  3.654 &  0.316 &  0.729 &  -15.627 &  2.638 &  0.827 &  0.344 \\
0.200 & 1535.0 &  1887.0 &  0.0694 &   -4.291 &  3.955 &  0.254 &  0.595 &  -10.485 &  2.692 &  0.618 &  0.387 \\
0.200 & 1575.0 &  1906.8 &  0.0677 &    3.464 &  4.031 &  0.439 &  0.637 &  -10.308 &  2.640 &  0.540 &  0.265 \\
0.200 & 1615.0 &  1926.4 &  0.0660 &    8.330 &  3.991 &  0.567 &  0.499 &  -10.810 &  2.576 &  0.524 &  0.186 \\
0.200 & 1655.0 &  1945.8 &  0.0644 &    3.729 &  4.024 &  0.344 &  0.748 &  -10.536 &  2.561 &  0.545 &  0.262 \\
0.200 & 1695.0 &  1965.0 &  0.0629 &    6.435 &  4.111 &  0.493 &  1.400 &  -12.606 &  2.557 &  0.635 &  0.403 \\
0.200 & 1735.0 &  1984.0 &  0.0614 &   -0.882 &  4.177 &  0.696 &  2.282 &   -9.990 &  2.533 &  0.611 &  0.659 \\
0.200 & 1775.0 &  2002.8 &  0.0600 &   -8.126 &  4.105 &  0.369 &  1.713 &   -8.385 &  2.429 &  0.567 &  0.547 \\
0.200 & 1815.0 &  2021.5 &  0.0587 &    0.684 &  4.115 &  0.317 &  0.175 &  -11.252 &  2.415 &  0.613 &  0.357 \\
0.200 & 1855.0 &  2039.9 &  0.0575 &    4.958 &  4.019 &  0.557 &  0.802 &  -12.005 &  2.325 &  0.633 &  0.385 \\
0.200 & 1895.0 &  2058.2 &  0.0562 &    3.502 &  3.838 &  0.377 &  1.037 &  -13.349 &  2.130 &  0.708 &  0.566 \\
0.200 & 1935.0 &  2076.4 &  0.0551 &    0.067 &  3.829 &  0.229 &  0.877 &  -11.955 &  2.073 &  0.659 &  0.348 \\
0.200 & 1975.0 &  2094.4 &  0.0540 &    1.574 &  3.852 &  0.309 &  0.829 &  -12.275 &  2.045 &  0.673 &  0.556 \\
0.200 & 2015.0 &  2112.2 &  0.0529 &   -3.076 &  3.872 &  0.163 &  0.606 &   -9.834 &  2.166 &  0.562 &  0.458 \\
0.200 & 2055.0 &  2129.9 &  0.0519 &   -0.166 &  3.756 &  0.144 &  0.447 &  -10.504 &  2.661 &  0.580 &  0.251 \\
0.200 & 2095.0 &  2147.5 &  0.0509 &    0.521 &  3.764 &  0.190 &  0.628 &  -12.039 &  2.622 &  0.655 &  0.304 \\
0.200 & 2135.0 &  2164.9 &  0.0499 &   -0.129 &  3.769 &  0.174 &  0.766 &  -12.558 &  2.555 &  0.687 &  0.374 \\
0.200 & 2175.0 &  2182.2 &  0.0490 &   -2.033 &  4.404 &  0.130 &  0.701 &  -10.399 &  2.292 &  0.585 &  0.248 \\
\end{tabular}
}}
\end{center}
\end{table*}
\pagebreak
\begin{table*}
\begin{center}
{\footnotesize{
\begin{tabular}{|c|c|c|c|c|c|c|c|c|c|c|c|}
0.200 & 2215.0 &  2199.3 &  0.0481 &    3.207 &  4.469 &  ~~~  0.320 ~~~  &  ~~~  0.843 ~~~  &  -10.564 & ~ 2.364 ~ &  ~~~  0.558  ~~~ & ~ 0.220 ~ \\
0.200 & 2255.0 &  2216.3 &  0.0473 &   -1.724 &  4.238 &  0.166 &  1.127 &   -8.273 &  2.255 &  0.465 &  0.276 \\
0.200 & 2295.0 &  2233.2 &  0.0464 &   -4.587 &  3.704 &  0.186 &  0.530 &   -8.809 &  2.296 &  0.508 &  0.191 \\
0.200 & 2335.0 &  2249.9 &  0.0456 &   -0.694 &  3.681 &  0.140 &  0.571 &   -8.788 &  2.071 &  0.722 &  0.258 \\
0.200 & 2375.0 &  2266.5 &  0.0449 &   -1.545 &  4.021 &  0.248 &  0.394 &   -7.649 &  2.309 &  1.340 &  0.537 \\
0.200 & 2415.0 &  2283.0 &  0.0441 &    1.542 &  4.321 &  0.191 &  0.822 &  -10.287 &  2.297 &  0.533 &  0.464 \\
0.200 & 2455.0 &  2299.4 &  0.0434 &    0.391 &  5.027 &  0.110 &  0.113 &   -8.194 &  2.753 &  0.445 &  0.151 \\
0.200 & 2495.0 &  2315.7 &  0.0427 &    3.748 &  4.711 &  0.295 &  0.214 &   -9.663 &  2.251 &  0.516 &  0.150 \\
0.200 & 2535.0 &  2331.8 &  0.0420 &   -3.462 &  5.315 &  0.176 &  0.098 &   -8.667 &  2.675 &  0.512 &  0.156 \\
0.200 & 2575.0 &  2347.9 &  0.0414 &   -4.257 &  5.050 &  0.185 &  0.200 &   -8.722 &  2.502 &  0.549 &  0.173 \\
0.200 & 2615.0 &  2363.8 &  0.0408 &    8.723 &  4.601 &  0.583 &  0.728 &   -8.317 &  2.103 &  0.429 &  0.275 \\
0.200 & 2655.0 &  2379.6 &  0.0401 &    3.937 &  5.560 &  0.269 &  0.943 &   -9.021 &  2.691 &  0.494 &  0.336 \\
0.240 &  277.5 &  1077.6 &  0.4609 &   14.650 &  4.723 &  4.956 &  2.176 &  -58.830 &  9.270 & 11.761 &  6.767 \\
0.240 &  320.0 &  1113.9 &  0.3997 &   12.332 &  5.068 &  3.379 &  2.528 &  -70.081 &  8.721 &  6.884 &  5.335 \\
0.240 &  360.0 &  1147.1 &  0.3553 &    7.192 &  3.694 &  2.159 &  2.890 &  -88.335 &  5.967 &  3.823 &  4.975 \\
0.240 &  400.0 &  1179.4 &  0.3197 &    2.021 &  3.783 &  3.169 &  3.019 & -120.832 &  5.717 &  4.758 &  4.737 \\
0.240 &  440.0 &  1210.8 &  0.2907 &    4.501 &  3.909 &  3.769 &  3.340 & -137.414 &  5.594 &  5.342 &  4.805 \\
0.240 &  480.0 &  1241.4 &  0.2664 &    6.774 &  3.906 &  3.661 &  3.144 & -128.423 &  5.340 &  4.992 &  4.292 \\
0.240 &  520.0 &  1271.3 &  0.2460 &   14.674 &  3.778 &  2.474 &  1.586 &  -80.133 &  4.966 &  3.240 &  2.040 \\
0.240 &  560.0 &  1300.5 &  0.2284 &    7.420 &  4.538 &  1.556 &  0.935 &  -49.234 &  5.786 &  1.970 &  1.101 \\
0.240 &  600.0 &  1329.0 &  0.2132 &    7.277 &  5.292 &  1.062 &  0.343 &  -31.454 &  6.538 &  1.279 &  0.407 \\
0.240 &  640.0 &  1357.0 &  0.1998 &    7.134 &  4.791 &  0.877 &  0.302 &  -25.574 &  5.755 &  1.026 &  0.413 \\
0.240 &  680.0 &  1384.3 &  0.1881 &    5.395 &  4.392 &  0.649 &  0.732 &  -19.033 &  5.145 &  0.738 &  0.730 \\
0.240 &  720.0 &  1411.2 &  0.1776 &    5.324 &  4.210 &  0.398 &  0.796 &   -8.635 &  4.816 &  0.359 &  0.743 \\
0.240 &  760.0 &  1437.5 &  0.1683 &    5.075 &  3.738 &  0.433 &  0.466 &   -9.629 &  4.143 &  0.443 &  0.651 \\
0.240 &  800.0 &  1463.4 &  0.1599 &   11.109 &  3.709 &  0.816 &  0.378 &   -9.196 &  3.833 &  0.620 &  0.465 \\
\end{tabular}
}}
\end{center}
\end{table*}
\pagebreak
\begin{table*}
\begin{center}
{\footnotesize{
\begin{tabular}{|c|c|c|c|c|c|c|c|c|c|c|c|}
0.240 &  840.0 &  1488.8 &  0.1523 &    8.491 &  3.343 &  ~~~  0.508  ~~~ &  ~~~  0.572 ~~~  &   -3.113 & ~ 3.381 ~ &   ~~~ 0.434 ~~~  & ~ 0.539 ~ \\
0.240 &  880.0 &  1513.8 &  0.1453 &   -1.176 &  3.047 &  0.373 &  1.171 &   -0.789 &  3.030 &  0.312 &  0.940 \\
0.240 &  920.0 &  1538.4 &  0.1390 &    0.851 &  2.695 &  0.282 &  0.745 &   -6.161 &  2.672 &  0.281 &  0.580 \\
0.240 &  960.0 &  1562.6 &  0.1332 &    0.765 &  2.565 &  0.229 &  0.344 &   -7.973 &  2.506 &  0.365 &  0.296 \\
0.240 & 1000.0 &  1586.5 &  0.1279 &    2.504 &  2.841 &  0.333 &  0.620 &   -4.932 &  2.610 &  0.266 &  0.657 \\
0.240 & 1040.0 &  1610.0 &  0.1230 &    7.956 &  3.699 &  0.488 &  0.925 &   -5.591 &  3.240 &  0.376 &  0.993 \\
0.240 & 1080.0 &  1633.1 &  0.1184 &    8.457 &  3.690 &  0.610 &  1.337 &  -10.510 &  3.174 &  0.581 &  0.804 \\
0.240 & 1120.0 &  1655.9 &  0.1142 &    7.155 &  3.552 &  0.714 &  1.814 &   -8.929 &  2.983 &  0.686 &  0.918 \\
0.240 & 1160.0 &  1678.4 &  0.1103 &   -2.486 &  3.677 &  0.271 &  1.572 &   -4.796 &  3.045 &  0.353 &  0.780 \\
0.240 & 1200.0 &  1700.7 &  0.1066 &    3.012 &  4.048 &  0.286 &  0.734 &   -8.299 &  3.277 &  0.380 &  0.531 \\
0.240 & 1240.0 &  1722.6 &  0.1031 &   -0.472 &  4.318 &  0.124 &  0.329 &   -4.527 &  3.370 &  0.244 &  0.461 \\
0.240 & 1280.0 &  1744.2 &  0.0999 &   -2.042 &  4.252 &  0.242 &  0.593 &   -8.851 &  3.259 &  0.499 &  0.351 \\
0.240 & 1320.0 &  1765.6 &  0.0969 &    5.330 &  4.242 &  0.478 &  0.522 &  -11.534 &  3.199 &  0.539 &  0.244 \\
0.240 & 1360.0 &  1786.7 &  0.0940 &    1.181 &  4.249 &  0.345 &  0.447 &  -12.226 &  3.152 &  0.633 &  0.312 \\
0.240 & 1400.0 &  1807.6 &  0.0914 &   10.185 &  4.394 &  0.741 &  1.519 &  -13.575 &  3.332 &  0.628 &  0.599 \\
0.240 & 1440.0 &  1828.3 &  0.0888 &    1.459 &  4.292 &  0.188 &  0.128 &   -7.033 &  3.187 &  0.365 &  0.150 \\
0.240 & 1480.0 &  1848.7 &  0.0864 &    2.829 &  4.054 &  0.323 &  0.300 &   -8.625 &  2.967 &  0.439 &  0.308 \\
0.240 & 1520.0 &  1868.9 &  0.0841 &   -1.767 &  3.848 &  0.250 &  0.831 &  -13.948 &  2.755 &  0.781 &  1.197 \\
0.240 & 1560.0 &  1888.8 &  0.0820 &   -4.030 &  4.057 &  0.270 &  0.714 &   -9.832 &  2.685 &  0.603 &  0.357 \\
0.240 & 1600.0 &  1908.6 &  0.0799 &    5.364 &  3.852 &  0.389 &  1.677 &   -9.721 &  2.455 &  0.478 &  1.420 \\
0.240 & 1640.0 &  1928.2 &  0.0780 &    6.050 &  3.600 &  0.506 &  1.492 &  -10.500 &  2.293 &  0.543 &  0.503 \\
0.240 & 1680.0 &  1947.5 &  0.0761 &    3.945 &  3.470 &  0.354 &  0.943 &  -10.827 &  2.190 &  0.613 &  0.329 \\
0.240 & 1720.0 &  1966.7 &  0.0744 &    3.671 &  3.498 &  0.356 &  1.263 &  -10.053 &  2.174 &  0.515 &  0.443 \\
0.240 & 1760.0 &  1985.7 &  0.0727 &   -0.841 &  3.164 &  0.523 &  1.685 &   -8.908 &  1.963 &  0.560 &  0.538 \\
0.240 & 1800.0 &  2004.5 &  0.0711 &   -2.307 &  3.086 &  0.324 &  1.077 &   -8.801 &  1.749 &  0.512 &  0.415 \\
0.240 & 1840.0 &  2023.2 &  0.0695 &    3.159 &  3.330 &  0.376 &  0.155 &  -10.280 &  2.061 &  0.531 &  0.255 \\
\end{tabular}
}}
\end{center}
\end{table*}
\pagebreak
\begin{table*}
\begin{center}
{\footnotesize{
\begin{tabular}{|c|c|c|c|c|c|c|c|c|c|c|c|}
0.240 & 1880.0 &  2041.6 &  0.0680 &    2.450 &  3.275 &  ~~~  0.338 ~~~  &  ~~~  0.899  ~~~ &  -10.236 & ~ 2.031 ~&  ~~~  0.549  ~~~ & ~ 0.319 ~ \\
0.240 & 1920.0 &  2059.9 &  0.0666 &    0.830 &  3.171 &  0.200 &  0.731 &   -9.732 &  1.944 &  0.521 &  0.482 \\
0.240 & 1960.0 &  2078.1 &  0.0653 &    3.365 &  3.127 &  0.333 &  0.535 &  -11.821 &  1.916 &  0.607 &  0.291 \\
0.240 & 2000.0 &  2096.1 &  0.0639 &    0.102 &  2.949 &  0.204 &  0.575 &  -11.153 &  1.617 &  0.600 &  0.391 \\
0.240 & 2040.0 &  2113.9 &  0.0627 &   -1.216 &  2.907 &  0.134 &  0.449 &   -9.477 &  1.765 &  0.520 &  0.254 \\
0.240 & 2080.0 &  2131.6 &  0.0615 &    1.361 &  3.078 &  0.192 &  0.110 &   -8.447 &  1.934 &  0.440 &  0.203 \\
0.240 & 2120.0 &  2149.1 &  0.0603 &    3.651 &  3.059 &  0.350 &  0.350 &  -13.124 &  1.902 &  0.668 &  0.527 \\
0.240 & 2160.0 &  2166.5 &  0.0592 &    1.622 &  3.051 &  0.244 &  0.600 &  -12.831 &  1.743 &  0.668 &  0.624 \\
0.240 & 2200.0 &  2183.8 &  0.0581 &   -0.408 &  3.305 &  0.128 &  0.444 &  -10.021 &  1.960 &  0.535 &  0.308 \\
0.240 & 2240.0 &  2200.9 &  0.0571 &   -1.516 &  3.507 &  0.153 &  0.317 &   -8.893 &  2.203 &  0.486 &  0.206 \\
0.240 & 2280.0 &  2217.8 &  0.0561 &   -5.569 &  3.579 &  0.245 &  0.334 &   -6.705 &  2.221 &  0.405 &  0.214 \\
0.240 & 2320.0 &  2234.7 &  0.0551 &   -4.849 &  3.695 &  0.278 &  0.720 &   -7.365 &  2.189 &  0.426 &  0.198 \\
 \hline 
 \end{tabular}
}}
\caption{Table of  $\sigma_\mathrm{LT}(\nu,Q^2)$ and $\sigma_\mathrm{TT}(\nu,Q^2)$ on $^3$He. From left to right: 
Four-momentum transfer squared; energy transfer $\nu$; Invariant mass $W$; 
Bjorken scaling variable $x$; 
Cross-section, statistical uncertainty, uncorrelated systematic uncertainty, and correlated uncertainty for $\sigma_\mathrm{LT}$ and $\sigma_\mathrm{TT}$, respectively.}
\label{tab:sigLT}

\end{center}
\end{table*}

\end{document}